\documentclass[12pt,letterpaper]{article}

\usepackage[affil-it]{authblk}
\usepackage{boxedminipage}
\usepackage{placeins}
\usepackage{geometry}
\usepackage[pdftex]{graphicx}
\usepackage{amssymb,amsfonts,amsmath}
\usepackage{color}
\usepackage{natbib}
\usepackage{hyperref}
\usepackage{amsthm,mathrsfs}
\usepackage[normalem]{ulem}
\usepackage{xr}
\newtheorem{theorem}{Theorem}
\newtheorem{corollary}{Corollary}

\newtheorem{lemma}{Lemma}

\newtheorem{definition}{Definition}

\newcommand{\argmin}{\operatornamewithlimits{arg\,min}}

\newcommand{\bv}{\mathbf{v}}

\newcommand{\A}{\mathbf{A}}
\newcommand{\B}{\mathbf{B}}
\newcommand{\C}{\mathbf{C}}
\newcommand{\D}{\mathbf{D}}
\newcommand{\mC}{\mathcal{C}}

\newcommand{\G}{\mathbf{G}}
\newcommand{\mG}{\mathcal{G}}

\newcommand{\X}{\mathbf{X}}

\newcommand{\bP}{\mathbf{P}}

\newcommand{\pa}{\texttt{pa}}
\newcommand{\bS}{\mathbf{S}}

\newcommand{\sgn}{\textrm{sgn}}
\newcommand{\adj}{\texttt{adj}}
\newcommand{\ch}{\texttt{ch}}
\newcommand{\de}{\texttt{de}}

\newcommand{\perpD}{\perp\!\!\!\perp}

\newcommand{\btheta}{\boldsymbol{\theta}}
\newcommand{\bXi}{\boldsymbol{\Xi}}
\newcommand{\bxi}{\boldsymbol{\xi}}
\newcommand{\bmu}{\boldsymbol{\mu}}
\newcommand{\bSigma}{\boldsymbol{\Sigma}}
\newcommand{\bOmega}{\boldsymbol{\Omega}}

\newcommand{\Con}{\texttt{Con}}

\newcommand{\vb}{\boldsymbol{b}}

\newcommand{\vt}{\boldsymbol{t}}
\newcommand{\vu}{\boldsymbol{u}}
\newcommand{\vx}{\boldsymbol{x}}
\newcommand{\vX}{\boldsymbol{X}}

\newcommand{\vzero}{\boldsymbol{0}}

\newcommand{\bbeta}{\boldsymbol\beta}
\newcommand{\beps}{\boldsymbol\epsilon}
\newcommand{\Kr}{\mathcal{K}}
\newcommand{\Er}{\mathcal{E}}
\newcommand{\Nr}{\mathcal{N}}
\newcommand{\Sr}{\mathcal{S}}
\newcommand{\vXcr}{\boldsymbol{\mathscr{X}}}
\newcommand{\trans}{\mathrm{T}}

\newcommand{\PP}{\mathbb{P}}
\newcommand{\RR}{\mathbb{R}}

\begin{document}
\title{\texttt{PenPC}: A Two-step Approach to Estimate the Skeletons of High Dimensional Directed Acyclic Graphs}

\author{Min Jin Ha\,$^1$, Wei Sun\ $^{1,2}$\footnote{to whom correspondence should be addressed}, and Jichun Xie\ $^{3}$}

\affil{$^{1}$Department of Biostatistics, University of North Carolina at Chapel Hill\\
       $^{2}$Department of Genetics, University of North Carolina at Chapel Hill\\
       $^{3}$Department of Statistics, Temple University }

\date{}
\maketitle

\begin{abstract}
 Estimation of the skeleton of a directed acyclic graph (DAG) is of
  great importance for understanding the underlying DAG and causal
  effects can be assessed from the skeleton when the DAG is not
  identifiable. We propose a novel method named \texttt{PenPC} to
  estimate the skeleton of a high-dimensional DAG by a two-step
  approach. We first estimate the non-zero entries of a concentration
  matrix using penalized regression, and then fix the difference
  between the concentration matrix and the skeleton by evaluating a
  set of conditional independence hypotheses. For high dimensional
  problems where the number of vertices $p$ is in polynomial or
  exponential scale of sample size $n$, we study the asymptotic
  property of \texttt{PenPC} on two types of graphs: traditional
  random graphs where all the vertices have the same expected number
  of neighbors, and scale-free graphs where a few vertices may have a
  large number of neighbors. As illustrated by extensive simulations
  and applications on gene expression data of cancer patients, \texttt{PenPC}
  has higher sensitivity and specificity than the standard-of-the-art
  method, the PC-stable algorithm.
\end{abstract}


\section{Introduction}

To understand the molecular mechanisms of human disease, huge amount
of high-dimensional genomic data have been collected from large number
of samples. For example, as of Feb 6th, 2014, The Cancer Genome Atlas
(TCGA) project \citep{mclendon2008comprehensive} has published
multiple types of genomic data in 8,909 cancer patients of 28
cancers. Many statistical methods have been developed to identify the
associations between genomic features and disease outcomes or cancer
subtypes. However, such association results are descriptive in their
nature, and they cannot deliver ``actionable'' conclusions for cancer
treatment. Many recently developed cancer drugs are so-called
``targeted drugs'' that target particular (mutated) proteins in cancer
cells, and the mechanism of such drugs can be understood as direct
interventions on tumor cells \citep{vogelstein2013cancer}. To
characterize or predict the consequences such drug interventions,
statical methods that allow causal inference based on high dimensional
genomic data are urgently needed.

One of the most commonly used tools for causal inference among a large
number of random variables is the directed acyclic graph (DAG) (also
known as Bayesian Network)
\citep{lauritzen1996graphical,pearl2009causality}. In a DAG, all the
edges are directed, and the direction of an edge implies a direct
causal relation. There is no loop in a DAG. Such ``acyclic'' property
is necessary to study causal relations
\citep{spirtes2000causation}. When we remove the directions of all the
edges in a DAG, the resulting undirected graph is the
\textit{skeleton} of the DAG.

Estimation of the skeleton of a DAG is of great importance. First, it
is a crucial step towards estimating the underlying DAG.  Second,
in many real data analyses where only observational data (instead of
interventional data) are available, the DAG is not identifiable but
the skeleton can be estimated; and previous studies have shown that
causal effects can be assessed from the skeleton of a DAG because a
limited number of edges of a DAG skeleton can be oriented using a set
of deterministic rules
\citep{maathuis2009estimating,maathuis2010predicting}. Several methods
have been developed to estimate DAGs or their skeletons from
observational data
\citep{heckerman1995learning, spirtes2000causation,
  chickering2003optimal, kalisch2007estimating},
however most of them are not suitable (theoretically and/or
computationally) for the high dimensional genomic problems that
motivate our study. For example, in the real data analysis presented
in Section 6, we sought to construct DAG of thousands of genes using
hundreds of samples.  In this paper, we proposed a new method named
\texttt{PenPC} to address this challenging problem. We proved the
estimation consistency of \texttt{PenPC} for high dimensional settings
of $p=O\left(\exp\{n^a\}\right)$ for $0 \leq a < 1$, and we also derived
the conditions for estimation consistency for two types of graphs:
random graph where all the vertices have the same expected number of
neighbors, and scale-free graphs where a few vertices can have much
larger number of neighbors than other vertices. As verified by both
simulation and real data analyses using TCGA data, \texttt{PenPC}
provides more accurate estimates of DAG skeletons than existing
methods.

The remaining parts of this paper are organized as follows.  In
Section 2, we give a brief review of DAG estimation methods and the
conceptual advantages of our \texttt{PenPC} algorithm. Details of the
\texttt{PenPC} algorithm are introduced in Section 3 and its
theoretical properties are presented in Section 4. We study the
empirical performance of \texttt{PenPC} and existing methods in
simulations and real data analyses in Section 5 and Section 6,
respectively. Finally, we conclude in Section 7.
 
\section{Review of DAG Estimation}

\subsection{Directed Acyclic Graph (DAG)}
A DAG of random variables $X_1, ..., X_p$ is a directed graph with no
cycle. Specifically, a DAG can be denoted by $\mG = (V, E)$,
where $V$ contains $p$ vertices $1, 2, ...., p$ that correspond to
$X_1,...,X_p$, and $E$ contains all the directed edges. In a DAG, a
\textit{chain} of length $n$ from $i$ to $j$ is a sequence $i = i_0 -
i_1 - \cdots - i_{n-1} - i_n = j$ of distinct vertices such that
$i_{l-1} \rightarrow i_l \in E$ or $i_{l} \rightarrow i_{l-1} \in E$
for $l=1,\ldots,n$; and a \emph{path} of length $n$ from $i$ to $j$ is
a sequence $i=i_0 \rightarrow i_1 \rightarrow \cdots \rightarrow
i_n=j$ of distinct vertices such that $i_{l-1} \rightarrow i_{l} \in
E$ for $l=1,...,n$. Given this path, $i_{l-1}$ is a \emph{parent} of
$i_{l}$, $i_{l}$ is a \emph{child} of $i_{l-1}$, $i_0, i_1, ...,
i_{l-1}$ are \emph{ancestors} of $i_l$, and $i_{l+1}, ..., i_n$ are
\emph{descendants} of $i_l$.

Given a DAG $\mG$ for random variables $X_1,\ldots,X_p$ and assume
that
\begin{equation} \X = (X_1,\ldots,X_p)^\trans \in \RR^p \sim P_X
  \textrm{ with density } f_X.\end{equation} We say that the
distribution $P_X$ is \emph{Markov} to $\mG$ if the joint density
$f_X$ satisfies the \textit{recursive factorization}
\begin{equation}\label{factDAG}
f(x_1,\ldots, x_p) = \prod_{i=1}^p f(x_i|x_{\pa_i}),
\end{equation}
where $\pa_i$ denotes the parents of vertex $i$. The factorization
naturally implies acyclic restriction of the graph
structure. Equivalently $P_X$ is Markov to $\mG$ if every variable is
conditionally independent of its non-descendants given its parents. A
related concept is the so-called \emph{faithfulness}:

\begin{definition}\label{def:faithfulness}
  Let $P_X$ be Markov to $\mG$. $<\mG,P_X>$ satisfies the faithfulness
  condtiontion if and only if every conditional independence relation
  true in $P_X$ is entailed by the Markov property applied to $\mG$
  \citep{spirtes2000causation}.
\end{definition}

This means that if a distribution $P_X$ is faithful to DAG $\mG$, all
conditional independences can be read off from the DAG $\mG$ using
d-separation defined in the following definition \ref{def:dsep1}, and
thus the faithfulness assumption requires stronger relationship
between the distribution $P_X$ and the DAG $\mG$ than the Markov
property.

\begin{definition}\label{def:dsep1} (d-separation). A vertex set $\bS$
  block a chain $\texttt{p}$ if either (i) $\texttt{p}$ contains at
  least one arrow-emitting vertex belonging to $\bS$, or (ii)
  $\texttt{p}$ contains at least one collision vertex (e.g., $j$ is a
  collision vertex if the chain includes $i \rightarrow j \leftarrow
  k$) that is outside $\bS$ and no descendant of the collision vertex
  belongs to $\bS$. If $\bS$ blocks all the chains between two sets of
  random variables $X$ and $Y$, we say ``$\bS$ d-separates $X$ and
  $Y$'' \citep{pearl2009causality}.
\end{definition}

Not all the distributions can be faithfully represented by a DAG. In
this paper, we assume the random variables follow multivariate normal
distribution, then the faithfulness assumption can be justified by the
fact that among all the multivariate normal distributions associated
with $\mG$, the non-faithful ones form a Lebesgue null set
\citep{meek1995strong}.

Given multivariate normal distribution assumption, a commonly used
graphical model is Gaussian Graphic Model (GGM), where two vertices
are connected if the corresponding two variables are independent,
given all the other variables. A GGM can be constructed by 
a \emph{concentration matrix} (i.e.,  precision matrix, or
inverse of covariance matrix) in that two vertices are connected if
the corresponding elements in the concentration matrix is
non-zero. The skeleton of a DAG is different from its GGM because of
v-structures. In a \emph{v-structure} $X \rightarrow W \leftarrow Z$,
co-parent $X$ and $Z$ are marginally independent or conditionally independent
given their parents, but given every vertex set that contains $W$ (a
collision vertex) or any descendant of $W$, $X$ and $Z$ are dependent
with each other. A few examples are shown in Figure 1, and instances
of the covariance and concentration matrices of the GGM in Figure 1(a)
are shown in the Supplementary Materials, Section 1.
\begin{figure}[h!]
  \centering
    \includegraphics[width=3.6in]{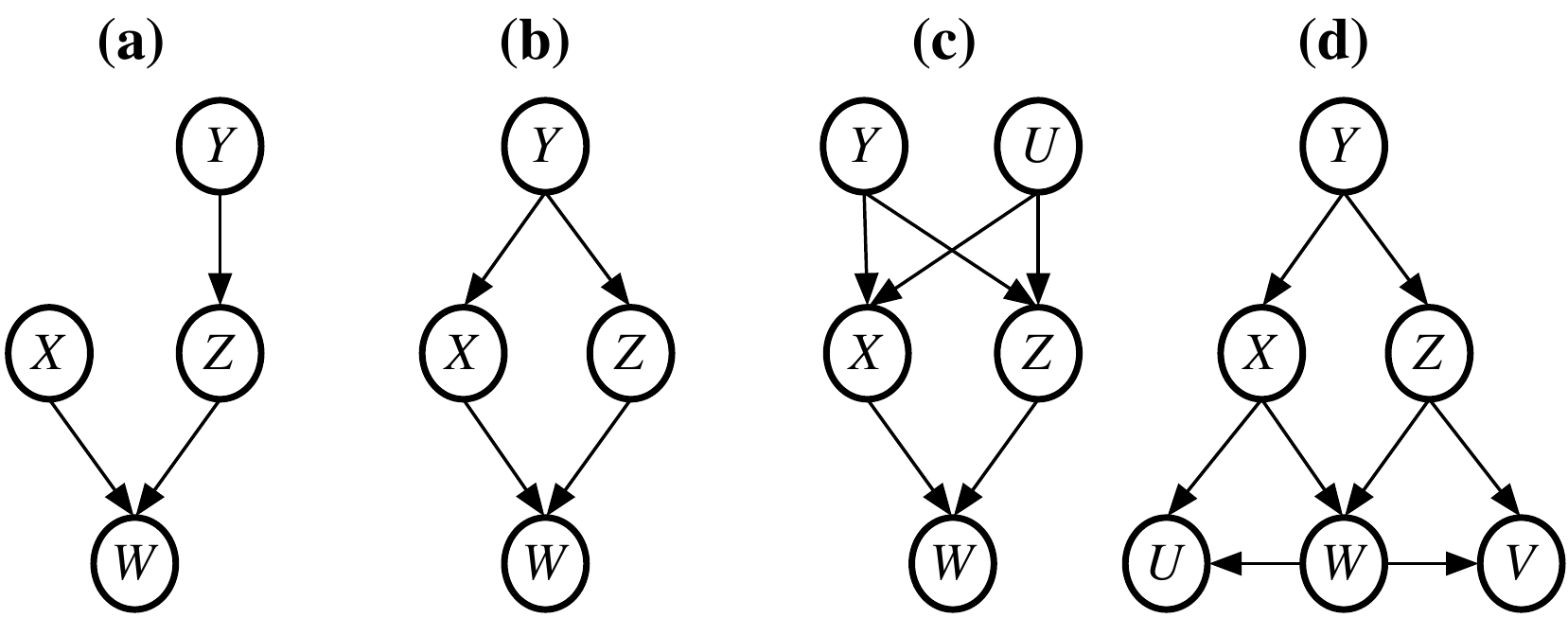}
    \caption{Four DAGs where $X$ and $Z$ are not connected in the
      skeleton, but are connected in the corresponding GGMs. The true
      relation between $X$ and $Z$ can be revealed by appropriate
      conditional independence testing. For example, $X \perp Z$ in
      Figure 1(a), $X \perp Z | Y$ in Figure 1(b), $X \perp Z | (Y,
      U)$ in Figure 1(c), and $X \perp Z | Y$ in Figure 1(d).  }
\end{figure}

\subsection{DAG estimation using observational data}
Many methods have been developed for DAG estimation using
interventional data. Since the focus of this paper is DAG skeleton
estimation using observational data, we will only provide a brief
review for relevant methods using observational data.

When the $p$ variables have a nature ordering (i.e., for any vertex
$X_i$, all the parents or ancestors of $X_i$ are among the vertices
$X_1$, ..., $X_{i-1}$, and all the children or descendants of $X_i$
are among vertices $X_{i+1}$, ..., $X_p$), the problem of skeleton
estimation is greatly simplified because a regression of $X_i$ versus
$X_1$, ..., $X_{i-1}$ can be used to identify the true skeleton
\citep{shojaie2010penalized}. However, in many high-dimensional
problems, such a nature ordering is not available. Throughout this
paper, we assume no knowledge of nature ordering.  Then the underlying
DAG is not identifiable from observational data, because conditional
dependencies implied by the Markov property on the observational
distribution $P_X$ only determine the \textit{skeleton} and
\textit{v-structures} of the graph \citep{pearl2009causality}. All the
DAGs with the same skeleton and v-structures correspond to the same
probability distribution and they form a \emph{Markov equivalence 
class}. After estimating skeleton, the v-structures can be identified by 
a set of deterministic rules, and thus we do not distinguish the 
estimation of a DAG skeleton and a Markov equivalence class. 

In general, there are two approaches for DAG or DAG skeleton estimation. The first one
is the search-and-score approach that searches for the DAG that
maximizes or minimizes a pre-defined score, such as BIC (Bayesian
Information Criterion). The second
one is the constraint-based approach that constructs DAGs by assessing
conditional independence of random variables. There are also some
hybrid methods that combine more than one method.

Direct search across all possible graphs is computationally infeasible
even for moderate number of variables. Elegant methods have been
developed to search across Markov equivalence classes
\citep{chickering2003optimal} or the nature orderings of the variables
\citep{teyssier2005ordering}. The objective function of
search-and-score methods (e.g., BIC) may be considered a
$L_0$-penalized likelihood, and a recent work shows several
theoretical merits of $L_0$-penalized maximum likelihood estimates
\citep{van2013l0}. These methods, however, are still computationally
very challenging for genomic applications where the number of vertices
can be thousands and sample size ranges from tens to
thousands.

One representative method of the constraint-based approach is the PC
algorithm (named after the first names its authors, Peter Sprites and
Clark Glymour) \citep{spirtes2000causation}. Starting with a complete
undirected graph where any two vertices are connected with each other,
the PC algorithm first thins the complete graph by removing edges
between vertices that are marginally independent. Then it removes
edges by assessing conditional independence given one vertex, two
vertices, and so on. \cite{kalisch2007estimating} proved the uniform
consistency of the PC-algorithm in high-dimensional settings where
$p=O(n^a)$ for $a>0$. The results of the PC algorithm depend on the
order of the edges to be assessed. \cite{colombo2012modification}
proposed a modification of the PC algorithm that overcomes such order
dependency. This new method, named as PC-stable algorithm, can
substantially improve the performance of the PC algorithm. We consider
the PC-stable algorithm as the state-of-the-art method for high
dimensional problems, and we will compare our method with the
PC-stable algorithm.

The Independence Graph (IG) algorithm (Chapter 5.4.3 of \cite{spirtes2000causation}) 
modifies the PC algorithm by using a different initial graph. Instead of
starting with a complete undirected graph as the PC algorithm, the IG
algorithm starts from an undirected independence graph, where two
vertices are connected if the corresponding two variables are
conditionally dependent given all the other variables. In such an
independence graph (with the assumption of no estimation error), the
neighbors of a vertex $Y_j$ include its parents, children and
co-parents in the underlying DAG, which constitute the so-called
Markov blanket of $Y_j$ such that $Y_j$ is independent of all the
other vertices given its Markov blanket
\citep{aliferis2010local}. Under multivariate normal distribution
assumption, independence graph is a Gaussian Graphic Model (GGM), and
thus can be determined by the concentration matrix.

The Max-Min Hill-Climbing (MMHC) algorithm is a popular hybrid method
that combines search-and-sore approach and constraint-based approach
\citep{tsamardinos2006max}.  The MMHC first estimates the skeleton of
the DAG using a constraint-based method (the Max-Min part of the
algorithm), and then orient the edges using a search-and-score
technique (the Hill-Climbing part of the algorithm).
\cite{schmidt2007learning} proposed to replace the Max-Min part of the
MMHC algorithm by a penalized regression with $L_1$ penalty, which
identifies the Markov blanket of each vertex and improves the
performance of the MMHC algorithm. \cite{meinshausen2006high} studied
the theoretical property of Markov blanket selection using the Lasso ($L_1$)
penalty, and they referred to this procedure as neighborhood selection. They pointed
out that selection consistency of a variable $Y$'s Markov blanket,
dented by $MB_{Y}$, requires a so-called irrepresentable condition
\citep{zhao2006model} that there is weak correlation between the
variables within and outside $MB_{Y}$. This is
a strong condition and it generally does not hold for the genomic
problems that motivate this study.

We propose a \texttt{PenPC} algorithm for DAG skeleton estimation in
two steps. It first adapts neighborhood selection method to select
Markov blanket of each vertex, and then it applies a modified
PC-stable algorithm to remove false positive edges due to
co-parents. Although the two-step approach of the \texttt{PenPC}
algorithm shares similar spirit to the IG algorithm
\citep{spirtes2000causation} and the modified MMHC algorithm
\citep{schmidt2007learning}, we have made the following novel
contributions. First, we employ the log penalty $p(|b|;\lambda,\tau) =
\lambda \log(|b| + \tau)$, one of the folded concave penalties
\citep{fan2011nonconcave}, for the neighborhood selection step, which
significantly improves the accuracy of Markov blanket search and allows
much stronger correlation between the variables within and outside 
a Markov blanket than what is allowed for the Lasso penalty. Combining the
neighborhood selection with log penalty and a novel modified PC-stable
algorithm, the resulting \texttt{PenPC} algorithm outperforms the
state-of-the-art PC-stable algorithm in terms skeleton estimation
accuracy. In high dimensional setting, the PenPC algorithm also enjoys
some advantage in terms of computational efficiency. Second, we
provide theoretical justifications of the estimation consistency of
the \texttt{PenPC} algorithm in high dimensional settings where
$p=O\left(\exp\{n^a\}\right)$ for $0 \leq a < 1$. We also discuss the
implications for estimation consistency for two types of graphs:
traditional random graph model where all the vertexes have the same
expected number of connections, and scale-free graph where a few
vertices can have much larger number of neighbors than the other
vertices. Whereas random graph is often assumed in previous studies,
e.g., for the consistency of the PC algorithm
\citep{kalisch2007estimating}, scale-free graph is more frequently
observed in gene networks as well as many other applications
\citep{barabasi1999emergence}.

\section{Methods}
We adopt a multivariate normal distribution assumption: $\X =
(X_1,\ldots,X_p)^\trans \sim N(0, \Sigma)$. Let $\vX = (\vx_1, ...,
\vx_{p})$ be the $n\times p$ observed data matrix. Without loss of
generality, we assume each column $\vx_i$ ($1 \leq i \leq p$) has
been standardized to have mean 0 and $\vx_j^\trans\vx_j=n$. Our
\texttt{PenPC} algorithm proceeds in two steps: (1) neighborhood
selection, and (2) application of a modified PC-stable algorithm to
remove false connections.

\textbf{Step 1. (Neighborhood Selection)} We first select the
neighborhood of vertex $i$ by a penalized regression with $X_i$ as
response variable and all the other variables corresponding to
vertices $V\setminus \{i\}$ as covariates:
\begin{eqnarray}\label{eq:pen}
  \hat{\vb}_i 
  =\argmin_{\vb_i\in \mathbb{R}^{p-1}} \frac{1}{2} (\vx_i -
  \vX_{-i}\vb_i)^\trans (\vx_i-\vX_{-i}\vb_i) + n\sum_{j \neq i}
  p(|b_{i,j}|; \lambda_i,\tau_i).
\end{eqnarray}
where $\vX_{-i}$ is an $n\times (p-1)$ matrix for $n$
measurements of the remaining $p-1$ covariates, $\vb_i=(b_{i,1}, ...,
b_{i,i-1}, b_{i,i+1}, ..., b_{i,p})^\trans$ and
$p(|b_{i,j}|;\lambda_i,\tau_i)$ denotes a penalty function with tuning
parameters $\lambda_i$ and $\tau_i$.  We consider a class of folded concave penalty
functions satisfying the following condition:
\begin{itemize}
\item[] \textbf{Condition 1:} The penalty function $p(t;\lambda,\tau)$ is
  of the form $\lambda \rho(t;\tau)$, where $\rho(t;\tau)$ is increasing
  and concave in $t\in[0,\infty)$ given $\tau$ and has continuous
  derivative $\rho'(t;\tau)$ in terms of $t$ and with $ \rho'(0+;\tau) >
  0$.
\end{itemize}
This is a generalization of the Condition 1 in
\cite{fan2011nonconcave}. Specifically, we employ the log penalty
$p(|b|;\lambda,\tau) = \lambda \log(|b| + \tau)$, which has been
demonstrated to have good performance in high-dimensional genetic
studies \citep{sun2010genomewide}. We employed the implementation of
penalized regression with log penalty using 
coordinate descent algorithm \citep{sun2010genomewide}, and the two tuning parameters $\lambda$
and $\tau$ are selected by two-grid search to minimize extended BIC
\citep{chen2008extended}.  After $p$ penalized regressions for each of
the $p$ variables, we construct the GGM by adding an edge between
vertices $i$ and $j$ if $\hat{b}_{ij} \neq 0$ or $\hat{b}_{ji} \neq
0$.
   
\textbf{Step 2. (Modified PC-stable algorithm)} We apply a modified
PC-stable algorithm to remove the false edges due to co-parent
relationships. For each edge $i - j$, we first assess marginal
association between vertices $i$ and $j$.  If they remain dependent,
we use the following strategy to search for candidate separation
sets. Let
\begin{itemize}
\item $\A_{i,j} = \left[\adj(i,\mC_{\mG}) \bigcup \adj(j,\mC_{\mG})\right]
  \setminus \{i,j\}$, i.e., the union of the neighbors of $i$ and $j$,
  except $i$ or $j$ themselves. $\A$ is the Markov blanket of $i$ and
  $j$.
\item $\B_{i,j} = \left[\adj(i,\mC_{\mG}) \bigcap \adj(j,\mC_{\mG})\right]
  \setminus \{i,j\}$, i.e., the intersection of the neighbors of $i$
  and $j$, except $i$ or $j$ themselves.
\item $\C_{i,j} = \{k: k \in \A \ \bigcap \ (\B_{i,j}\ \bigcup \
  \Con_{\mC_{\mG}}^{(i,j)}(\B_{i,j})) \}$, where
  $\Con_{\mC_{\mG}}^{(i,j)}(\B_{i,j})$ is the set of vertices
  connected to any vertex in $\B_{i,j}$ by a chain of any length from
  a subgraph of $\mC_{\mG}$, which is created by removing vertices $i$
  and $j$ as well as any edges connected to $i$ or $j$. Obviously
  $\B_{i,j} \subseteq \C_{i,j}$.
  \end{itemize}

Then the candidate conditional sets are
 \begin{eqnarray}\label{eq:pi}
 \boldsymbol{\Pi}_{i,j} = \{\A  \setminus \D, \D \subseteq \C\}.
 \end{eqnarray}
 Note that in the definition of $\boldsymbol{\Pi}_{i,j}$, we skip the
 subscript $_{i,j}$ for $\A$, $\B$, $\C$, and $\D$ to simplify
 notations. Each element of $\boldsymbol{\Pi}_{i,j}$ is a set $\A
 \setminus \D$, where $\D$ is exhaustively searched across all subsets
 of $\C$. The number of candidate conditional sets are $2^{|\C|}$,
 which is often much smaller than all the conditional sets
 $2^{|\A|}$. More details are described in the Supplementary
 Materials, Section 2. An intuitive explanation is as follows. By
 Markov property in (\ref{factDAG}), the d-separation set of $i$ and
 $j$ consists of their parents, but not their shared children or
 descendants. All the parents of $i$ or $j$ belong to set $\A$. All
 the shared descents of $i$ and $j$ (among those within the the Markov
 blanket of $i$ and $j$) belong to $\C$. Therefore we define
 $\boldsymbol{\Pi}_{i,j}$ such that it iteratively excludes any set of
 vertices that are likely to be the shared children/descendants of $i$
 and $j$.

 We test the conditional independence of $X_i$ and $X_j$ given $\Kr
 \in \boldsymbol{\Pi}_{i,j}$ using Fisher transformation of partial
 correlation. Specifically, denote the partial correlation between
 $X_i$ and $X_j$ given $\Kr \in \boldsymbol{\Pi}_{i,j}$ by
 $\rho_{i,j|\Kr}$. With the significance level $\alpha$, we reject the
 null hypothesis $H_0:\rho_{i,j|\Kr} = 0$ against the alternative
 hypothesis $H_a:\rho_{i,j|\Kr}\neq 0$ if
 $\sqrt{n-|\Kr|-3}\hat{z}_{i,j|\Kr}>\Phi^{-1}(1-\alpha/2)$, where
 $\hat{z}_{i,j|\Kr}=0.5\log((1+\hat{\rho}_{i,j|\Kr})/(1-\hat{\rho}_{i,j|\Kr}))$
 and $\Phi(\cdot)$ is the cdf of $N(0,1)$.

 The final output of \texttt{PenPC} algorithm is the estimated
 skeleton and separation sets $S(i,j)$ for all $(i,j)$. The separate
 sets are needed for causal effect estimation. If vertices $i$ and $j$
 are not connected in the GGM (then they won't be connected in the
 skeleton), their separation set is all the remaining variables. If
 $i$ and $j$ are connected in both the GGM and the skeleton, there is
 no separation set. If $i$ and $j$ are connected in the GGM, but not
 the skeleton, the separation set $S(i,j)$ is a set belongs to
 $\boldsymbol{\Pi}_{i,j}$, such that the test $X_i \perp X_ j \mid
 S(i,j)$ gives affirmative conclusion. Given the skeleton and the
 separation sets, causal effects can be assessed using function
 \texttt{idaFast} of R package pcalg \citep{kalisch2012causal}.

\section{Theoretical Properties}
\subsection{Fixed Graphs}
We denote the $L_2$ and $L_\infty$ norm of a matrix/vector by
$\|\cdot\|_2$ and $\|\cdot\|_\infty$, respectively. The $L_2$ norm of
a symmetric matrix is the maximum eigenvalue of the matrix. The
$L_\infty$ norm of a matrix is the maximum of the $L_1$ norm of each
row. The $L_\infty$ norm of a vector is the maximum of the absolute
values of its elements.  In this section we study high dimensional
behavior where $p$ grows as a function of sample size $n$. Thus we
denote $p$ as $p_n$, and denote a DAG and the corresponding GGM by
$\mG_n=(V_n,E_n)$ and $\mC_{\mG_n}=(V_n,F_n)$, respectively. We
further denote the skeleton of $\mG_n$ by $\mG_n^u=(V_n,E_n^u)$ where
$a - b\in E_n^u \Leftrightarrow a \rightarrow b\in E_n$ or $b
\rightarrow a\in E_n$.  For any vertex $i$, denote the observed data
of the variables within and outside of $\adj(i,\mC_{\mG_n})$ (but not
including $\vx_i$) by $\vXcr_{i1}$ and $\vXcr_{i2}$, respectively.  If
the penalty function has continuous second derivative, we define
$\kappa(\bv;\lambda,\tau) = \max_{1\leq j\leq r} -\lambda
\rho''(\lvert v_j\rvert;\tau)$, where for $\bv=(v_1,...,v_r)^\trans\in
\mathbb{R}^{r}$ and $v_j\neq 0$; otherwise we replace $\rho''(\lvert
v_j\rvert;\tau)$ by $\lim_{\epsilon\rightarrow 0 +} \sup_{t_1<t_2\in
  (\lvert v_j\rvert - \epsilon,\lvert v_j \rvert + \epsilon )}
\frac{\rho'(t_2;\tau_i) - \rho'(t_1;\tau_i)}{t_2-t_1}.$

The following conditions are needed for the
consistency of the \texttt{PenPC} algorithm.
\begin{itemize}

\item[(A1)] Dimensionality of the problem. $p_n = O\left(\exp\{n^a\}\right)$ with
  $a\in[0,1)$. 
  
\item[(A2)] Sparseness assumption. Let $q_n = \max_{1\leq j \leq p_n} |\adj(j, \mC_{\mG_n})|$,
 i.e., the maximum degree of $\mC_{\mG_n}$. $q_n \leq
  O(n^{b})$ for some $0 \leq b < 1$. By the following
  Lemma~\ref{lm:alg}, $\max_{1\leq j \leq p_n} |\adj(j,\mG_n)| \leq
  q_n$.

\item[(A3)] Minimum effect size for neighborhood selection. $\delta_n \equiv
  (1/2)\inf_{i,j} \left\{\left| b_{i,j} \right|: b_{i,j} \neq 0 \right\} \geq O(n^{-d_2})$ with $0<d_2<(1-a)/2-s_0$, where $s_0$ is a constant describing the correlation structure of the covariates with non-zero effect: 
  $\|(\vXcr_{i1}^\trans \vXcr_{i1})^{-1}\|_{\infty} = O(n^{-1 +
    s_0})$ with $0 \leq s_0< (1-a)/2$. 

\item[(A4)] Conditions for penalty function. $p'(\delta_n; \lambda_i, \tau_i)  \ll n^{-d_2 -
    s_0}$, $p'(0+; \lambda_i, \tau_i) \gg n^{-1/2 + a/2+b } \sqrt{\log n}$.

\item[(A5)] Further conditions for penalty function with respect to covariance structure of 
the covariates. For all $i=1,\ldots,p$ and $0 < K < 1$, 
  $\left\|
    \vXcr_{i2}^\trans \vXcr_{i1} (\vXcr_{i1}^\trans \vXcr_{i1})^{-1}
  \right\|_\infty \leq \min(K \frac{\rho'(0+; \tau_i)}
  {\rho'(\delta_n;\tau_i)}, O(n^b))$ and $\lVert
  (\vXcr_{i1}^\trans\vXcr_{i1})^{-1} \rVert_2 < 1/(n\kappa_0),$
  where $\kappa_0 = \max_{\bbeta_1 \in\Nr_{i1}}
  \kappa(\bbeta_1;\lambda_i,\tau_i)$, and $\Nr_{i1}$ is a hypercube around $\vb_{i1}$ (a sub-vector of $\vb_i$'s non-zero components) such that $\|\bbeta_{1} - \vb_{i1}\|_\infty \leq Cn^{-d_2}$.

\item[(A6)] Restriction on the size of conditional partial correlation. Denote the partial correlations between $X_i$ and $X_j$
  given a set of variables $\{X_r: r \in \Kr\}$ for $\Kr \subseteq V_n
  \setminus \{i, j\}$ by $\varrho_{i, j|\Kr}$. 
  For $\Kr \in \boldsymbol{\Pi}_{ij}$, the absolute values of
  $\varrho_{i, j|\Kr}$'s are bounded:
 \[
    \inf_{i,j,\Kr}\left\{\left| \varrho_{i, j|\Kr} \right|: \rho_{i,
        j|\Kr} \neq 0, \Kr\in\boldsymbol{\Pi}_{i,j} \right\} \geq c_{n}, 
    \textrm{ and }
    \sup_{i,j,\Kr}\left\{ \left| \varrho_{i, j|\Kr} \right|:
    \Kr \in \boldsymbol{\Pi}_{ij}\right\} \leq M < 1,
  \]
  where $c_{n} = O(n^{-d_1})$ for some
  $0<d_1<\min\{(1-a)/2,(1-b)/2\}$.  
   \end{itemize}

The sparseness assumption (A2) will
be replaced by tighter assumptions for two specific random graph
models later. Assumptions (A3)-(A5) ensure
that the step 1 of \texttt{PenPC} can recover the GGM. 
Assumption (A6) ensures the summation of the mistaken probabilities 
of the step 2 of the \texttt{PenPC} algorithm goes to 0 asymptotically. 
The condition 
$\left\|\vXcr_{i2}^\trans \vXcr_{i1} (\vXcr_{i1}^\trans \vXcr_{i1})^{-1}
  \right\|_\infty \leq K {\rho'(0+; \tau_i)} / {\rho'(\delta_n;\tau_i)}$ 
  in Assumption (A5) deserves more discussion 
  since it reveals why our neighborhood 
selection method using log penalty can perform better than the Lasso. 
For the Lasso, there is no extra parameter $\tau_i$ and $\rho(t)=|t|$, and thus  
$\rho'(0+)/\rho'(\delta_n) = 1$. Therefore the condition becomes 
$\left\|\vXcr_{i2}^\trans \vXcr_{i1} (\vXcr_{i1}^\trans \vXcr_{i1})^{-1} \right\| \leq K$, which 
is equivalent to the irrepresentable condition. In contrast, for the log penalty, 
$\rho'(t;\tau_i)=\sgn(t)/(|t| + \tau_i)$, and thus 
$\rho'(0+; \tau_i)/\rho'(\delta_n; \tau_i) \rightarrow (\delta_n + \tau_i)/\tau_i$, which can goes to 
infinity if $\tau_i = o(\delta_n)$. More specifically, the scale of $\rho'(0+; \tau_i)/\rho'(\delta_n; \tau_i)$ can be derived as follows. By assumption A4, $\rho'(0+; \tau_i)/\rho'(\delta_n; \tau_i) \gg (n^{-1/2 + a/2+b} \sqrt{\log n}) / (n^{-d_2 - s_0}) = n^b\sqrt{\log n} \rightarrow \infty$, where the last equality is due to Assumption A3. We can show that the log penalty satisfies other assumptions and refer the readers to \cite{Chen2014} for details. 

The following Lemma~\ref{lm:reg} claims that the support of the regression
coefficients is the same as that of the concentration
matrix. Therefore, we can use the regression model to estimate the GGM
$\mC_{\mG_n}$.
\begin{lemma}\label{lm:reg}
  Suppose $X=(X_1,...,X_p)^\trans \sim \mathcal{N}_p(\bmu, \bSigma)$
  and $\bOmega=\bSigma^{-1}$.  Then
  \begin{equation}
  X_i = X_{-i}^\trans \vb_{i} + \epsilon_i,\label{eq:bi}
\end{equation}
where $X_{-i}$ denotes
  a random vector derived from $X$ by removing $X_i$ from $X$, $\vb_i
  =- \sigma_i^2 \bOmega_{-i,i}$, and $\epsilon_i \sim
  \mathrm{N}(0,\sigma_i^2)$, with $\sigma_i^2 = \bSigma_{ii} -
  \bSigma_{i,-i} (\bSigma_{-i,-i})^{-1}\bSigma_{-i,i}$. $\bSigma_{ab}$
  and $\bOmega_{ab}$ are the sub-matrices where the subscripts $a$ and
  $b$ indicate inclusion/exclusion of certain random variables.
\end{lemma}

The proof of Lemma~\ref{lm:reg} is omitted since it is straightforward conclusion based on
conditional distribution of multivariate normal random variables.

Consider the neighborhood selection problem for one of the variables
versus all the other variables. Let $\Sr_i = \textrm{supp}(\vb_i)$ be
the support of the true regression coefficient $\vb_i$ with size
$|\Sr_i| = s_i$. From Lemma~\ref{lm:reg}, the degree of vertex $i$ in
$\mC_{\mG_n}$ is $s_i$. Recall that in assumption (A4) $\vXcr_{i1}$
and $\vXcr_{i2}$ denote the observed data of the variables
corresponding to $\Sr_i\subseteq V_n\setminus\{i\}$ and its
complement, $\Sr_i^c = V_n\setminus \left(\Sr_i\bigcup
  \{i\}\right)$. Similarly $\vb_{i1}$ and $\hat{\vb}_{i1}$ are
respectively the sub-vectors of $\vb_i$ and $\hat{\vb}_i$
corresponding to $\Sr_i$.

\begin{theorem}\label{th:rec}
  Given Assumptions (A1) - (A5), with probability at least
  $1-C\exp\{n^a-n^a\log(n)\}$ for a constant $0<C<\infty$, there
  exists a local minimizer $\hat{\vb}_i =
  (\hat{\vb}_{i1},\hat{\vb}_{i2})^\trans$ that satisfies the following
  conditions: for any $i = 1,\ldots, p_n$,
\begin{enumerate}
  \item[(a)] Sparsity: $\hat{\vb}_{i2} = \vzero$.
  \item[(b)] $L_{\infty}$ loss: $\|\hat{\vb}_{i1} -
    \vb_{i1}\|_{\infty} = o(n^{-d_2})$, where $d_2$ is defined in (A3).
\end{enumerate}
 \end{theorem}
 The proof is in the Supplementary Materials.  Under assumption (A1),
 the dimensionality $p_n$ is allowed to grow up to exponentially fast
 with sample size $n$. The value of $d_2$ can be as large as $1/2$
 depending on the lower bound of effect size specified in assumption
 (A3).

 Corollary~\ref{co:prec} is a simple extension from
 Theorem~\ref{th:rec}. It characterizes the consistency of the $p_n$
 penalized regression models which estimate the GGM
 $\mC_{\mG_n}$. Denote $\hat{\mC}_{\mG_n}(\btheta)$ as the estimate of
 $\mC_{\mG_n}$ by the neighborhood selection, where $\btheta$ are
 tuning parameters of the penalty function.

\begin{corollary}\label{co:prec}
  Given Assumption (A1), (A4)-(A6),
  \[\PP( \hat{\mC}_{\mG_n}(\btheta) = \mC_{\mG_n} )\geq
  1-C\exp\{2n^a-n^a\log(n)\}\] for a constant $0<C<\infty$.
\end{corollary}

\begin{lemma}\label{lm:alg}
  If the distribution $P_X$ is \emph{Markov} to $\mG$, i.e., if the
  joint density $f_X$ satisfies the recursive factorization, the set
  of edges $F_n$ of $\mC_{\mG_n}$ includes all edges $E_n^u$ of
  $\mG_n^u$ plus co-parent relationship in $\mG_n$.
\end{lemma}
This lemma \ref{lm:alg} has been proved in Lemma 3.21 of
\cite{lauritzen1996graphical}.

\begin{lemma}\label{lm:pi}
  Assume (A1). If $(i,j)\in F_n$ of $\mC_{\mG_n}$ but $(i,j)\notin
  E_n^u$ of $\mG_n^u$, the conditioning set $\boldsymbol{\Pi}_{i,j}$
  in (\ref{eq:pi}) includes at least one set which d-separates
  vertices $i$ and $j$ in $\mG$.
\end{lemma} 

The proof of Lemma 3 is presented in the Supplementary
Materials. Lemma~\ref{lm:alg} and Lemma~\ref{lm:pi} provide the
theoretical justifications for using GGM as a starting point of our
modified PC-algorithm.  Lemma~\ref{lm:alg} shows that if we have a
perfect estimation of the concentration matrix, we can recover all the
edges in the skeleton with no false negatives, but some false
positives: the co-parent relationships. Lemma~\ref{lm:pi} presents
that we can remove the false positives due to co-parent relationship
by examining partial correlation conditioning on some set in
$\boldsymbol{\Pi}_{i,j}$.

Next we discuss the theoretical property of the modified PC-stable
algorithm (the second step of the \texttt{PenPC} algorithm) given a perfect estimation of GGM. Later we will show that the
summation of mistaken probabilities of GGM estimation and skeleton
estimation given GGM goes to 0 as $n \rightarrow \infty$.
\begin{theorem}\label{th:pc}  Let $\alpha_n$ be the p-value
threshold for testing whether a partial correlation is 0. Let $\hat{\mG}^u_n(\alpha_n)$ be the
  estimates of $\mG^u_n$ from the second step of the \texttt{PenPC}
  algorithm given a perfect estimation of GGM from the first step of the
  \texttt{PenPC} algorithm.  Assume (A1), (A2) and (A6), then there exists $\alpha_n
  \rightarrow 0$, such that
\begin{eqnarray*}
  \PP\left[\hat{\mG}^u_n(\alpha_n) = \mG^u_n\right] = 1 - O\left(\exp\{-C
    n^{1 - 2d_1}\}\right) \rightarrow 1,
\end{eqnarray*}
where $0 < C < \infty$ is a constant. 
\end{theorem}

The proof is in the Supplementary Materials. Similar theorem has been
proved in \cite{kalisch2007estimating} with $p_n$ at polynomial order
of $n$. By exploiting accurate estimation of GGM, we extend the theorem to
$p_n=O\left(\exp\{n^a\}\right)$ case. Corollary~\ref{co:comb} provides
the combined error of step 1 and step 2 of $\texttt{PenPC}$ algorithm
as a simple extension of Corollary~\ref{co:prec} and
Theorem~\ref{th:pc}. 
\begin{corollary}\label{co:comb}
  Let $\hat{\mG}^u_n(\btheta,\alpha_n)$ be the
estimates of $\mG^u_n$ from the two-step approach \texttt{PenPC}
algorithm.
Assume (A1)-(A6), then there exists an
  $\alpha_n\rightarrow 0$, such that
\begin{eqnarray*}
  \PP\left[\hat{\mG}^u_n(\btheta,\alpha_n) = \mG^u_n\right] = 1 - O\left(\exp\{-C
    n^{1 - 2d_1}\}\right) \rightarrow 1,
\end{eqnarray*}
where $0 < C < \infty$ is a constant.
\end{corollary}

\subsection{Random Graphs}

Under certain conditions, the theoretical results could also be
extended to two commonly used models for random graphs: Erd\H{o}s and
R\'{e}nyi (ER) Model \citep{erdos1960evolution} and Barab\'{a}si and
Albert (BA) Model \citep{barabasi1999emergence}. In general,
assumption (A2) no longer holds for random graphs. However, based on
the proof in the Supplementary Materials, it is easy to see that
assumption (A2) can be relaxed to (A2').

\begin{itemize}
\item[(A2')] Let  $q_n = \max_{1\leq j \leq p_n} |\adj(j,
  \mC_{\mG_n})|$. 
 Assume 
\[ \PP\{q_n \leq
 O(n^{b})\} = 1,\quad \text{for some}\  0 \leq b < 1. \]
\end{itemize}
It is then suffices to show assumption (A2') holds. Note that the
value of $b$ in this assumption will affect the minimum effect size of
partial correlations in assumption (A3) and the convergence
probability in Theorem~\ref{th:rec} and Corollary~\ref{co:prec}.

\subsubsection{Erd\H{o}s and R\'{e}nyi (ER) Model}

The ER model constructs a graph $G(p_n,p_E)$ of $p_n$ vertices by
connecting vertices randomly. Each edge is included in the graph with
probability $p_{E}$ independent from all other edges. By law of large
numbers, such vertex is almost surely connected to $(p_n-1)p_E$
edges. Let $M_n$ be the maximal degree of the
graph. \cite{erdos1960evolution} proved the following results about
$M_n$.
\begin{lemma}\label{lm:ERdegree}
  In the graph $G(p_n,p_E)$ following the ER model, the maximal degree
  $M_n$ almost surely converges to $m_n$, where
 \[m_n =
 \begin{cases}
   O(\log p_n), & \text{if}\  p_np_E < 1,\\
   p_n^{2/3}, & \text{if}\  p_np_E = 1,\\
   O(p_n), & \text{if}\  \lim_{p_n\rightarrow \infty} p_np_E = c>1.
 \end{cases}
\]
\end{lemma}
\noindent When $p_n = O\{\exp(n^a)\}$, by Lemma~\ref{lm:ERdegree}, assumption (A2') holds immediately if $p_np_E < 1$ and $b \geq a$. When $p_np_E\geq 1$, our proof cannot
handle the general case $p_n = O\{\exp(n^a)\}$. However, when the
number of vertices is of the polynomial order of $n$, assumption (A2')
may still hold. In particular, suppose $p_n= O(n^r)$. When $p_np_E <
1$, assumption (A2')  holds for any $b \in \ [0,\infty)$. When $p_np_E =
1$, assumption (A2') holds if $b \geq 2r/3$. When $p_np_E \rightarrow
c>1$, assumption (A2') holds if $r < 1$ and $b \geq r$.

\subsubsection{Barab\'{a}si and Albert (BA) Model}
The BA model is used to generate scale free graphs whose degree
distribution follows a power law: $\PP(\nu) =
\gamma_0\nu^{-\gamma_1},$ with a normalizing constant $\gamma_0$ and a
exponent $\gamma_1$.  Specifically, BA model generates a graph by
adding vertices into the graph over time and when each new vertex is
introduced into the graph, it is connected with larger probability to
the existing vertices with larger number of connections. Since the
distribution does not depend on the size of the network (or time), the
graph organizes itself into a scale free state
\citep{barabasi1999emergence}. \cite{mori2005maximum} showed that
$M_n$ (the maximal degree of the graph) almost surely converges to $O(p^{1/2})$. 
Thus, assumption (A2') holds for the case $p_n = O(n^r)$ with $b \leq r/2$.

\section{Simulation Studies}
%
We evaluated the performance of the \texttt{PenPC} algorithm and the PC-stable algorithm in terms of sensitivity and specificity of skeleton estimation using DAGs simulated by the ER model or the BA model. In both simulations and real data analysis, we used the implantation of the PC-stable algorithm  by function \texttt{skeleton} in \texttt{R} package \texttt{pcalg} (version 1.1-6), and we have implemented \texttt{PenPC} algorithm in \texttt{R} package \texttt{PenPC}. 

Following \cite{kalisch2007estimating}, we simulated DAGs of $p$
vertices by the ER model as follows. First we assumed the $p$ vertices
were ordered so that if $i < j$, vertex $i$ could only be the parent
rather than child of vertex $j$. Then for any vertex pair $(i,j)$
where $i < j$, we added an edge $i \rightarrow j$ with probability
$p_{E}$. For the BA model, the DAGs were simulated following
\cite{barabasi1999emergence}. The initial graph had one vertex and no
edge. Then a new vertex was added in each step and directed edges were
added so that they started from the new vertex and pointed to some of
the existing vertices. Specifically, in the $(t+1)$-th step, $e$ edges
were proposed. For each edge, the new vertex was connected to the
$i$-th ($1\leq i\leq t$) existing vertex with probability
$\nu_i^{(t)}/\sum_j \nu_j^{(t)}$, where $\nu_i^{(t)} =
|\adj(i,\mG^{(t)})|$, and $\mG^{(t)}$ was the DAG at the $t$-th step,
right before adding the new vertex. Figure 2 shows the distribution of the
degrees $\nu$ from simulated DAGs under ER model ($p=1000$ and
$p_E=2/p$) and BA model ($p=1000$ and $e=1$).
\begin{figure}[h!]
{ \centering
\includegraphics[width=5.5in]{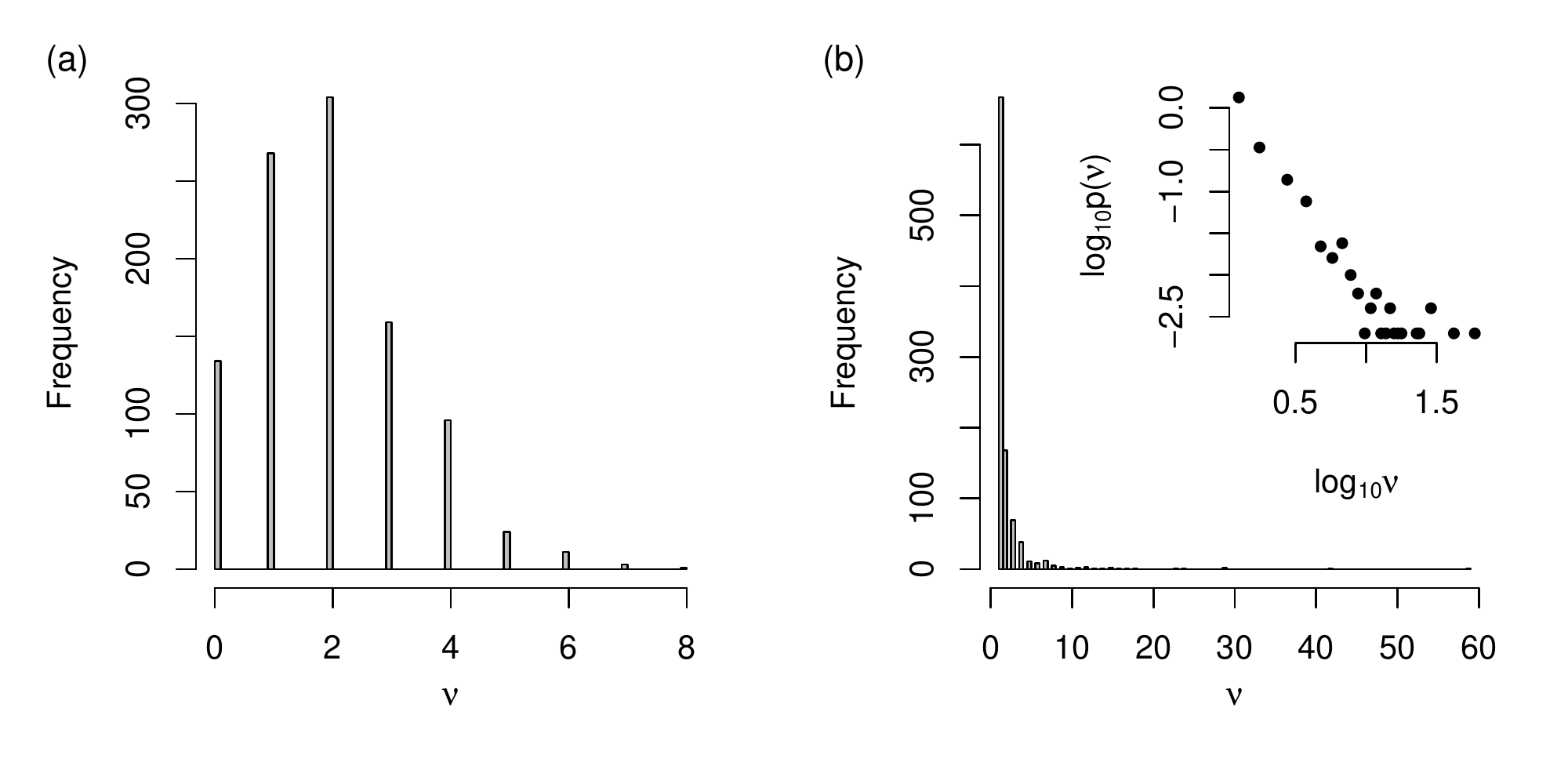}
}
\vspace{-2ex}
\caption{Histograms of the degree $\nu$. (a) ER model with $p=1000$
and $p_E=2/p$. (b) BA model with $p=1000$ and $e=1$ and the
$\log_{10}$ scale density of $\log_{10}\nu$ in its subplot.}
\end{figure}
The probability of finding a highly connected vertex decreases
exponentially with $\nu$ for the graphs generated by the ER model
(Figure 2(a)). However, for the graphs generated by the BA model,
highly connected vertices with large $\nu$ have relatively large
chance of occurring (Figure 2(b)), and there is a linear relation
between degree and degree probability in log-log scale, which confirms
the scale-free property of the graphs generated by the BA
model. Similar conclusions apply for the graphs generated by the BA
model with $e=2$ (Figure S1 of the Supplementary Materials).

After constructing the DAGs, the observed were are simulated by
structure equations under multivariate normal assumption. For example,
denote the parents of $X_j$ by $\pa_j$, then $\vx_j = \sum_{k\in
  \pa_j} b_{jk}\vx_k + \mathbf{\epsilon}_j$, where $
\mathbf{\epsilon}_j \sim \mathcal N(0, \sigma^2 I_{n\times n})$. In
our simulations, all $b_{jk}$'s and $\sigma^2$ were set to be 1. Our
simulation settings were displayed in Table 1.
\begin{table}[h!]
\small
   \caption{Simulation Setting}
 \centering
    \begin{tabular}{cccc}
    \hline
     $p$ & $n$ & $p_E$ (ER) & $e$ (BA)\\
    \hline
     11 & 100 & 0.2 & 1,2 \\
    100 & 30 & 0.02, 0.03, 0.04, 0.05 & 1,2\\
    1000 & 300 & 0.002, 0.005, 0.01 & 1,2 \\
    \hline
    \end {tabular}
     \label{tab.Params}
     \normalsize
\end{table}
For either ER or BA model, we considered low dimension setting where
$p=11, n=100$ and high-dimension settings where $p=100,n=30$ and
$p=1000,n=300$ with various sparsity levels determined by $P_E$ for ER
model and $e$ for BA model. Due to limited space, here we only show
the results for the simulation setups using ER model $(p=1000,n=300,
p_E=0.005)$ or BA model $(p=1000,n=300,e=1)$, and other results are
presented in Figure S3 - Figure S15 of the Supplementary Materials.
\begin{figure}[h!]
 \centering
\includegraphics[width=1\textwidth]{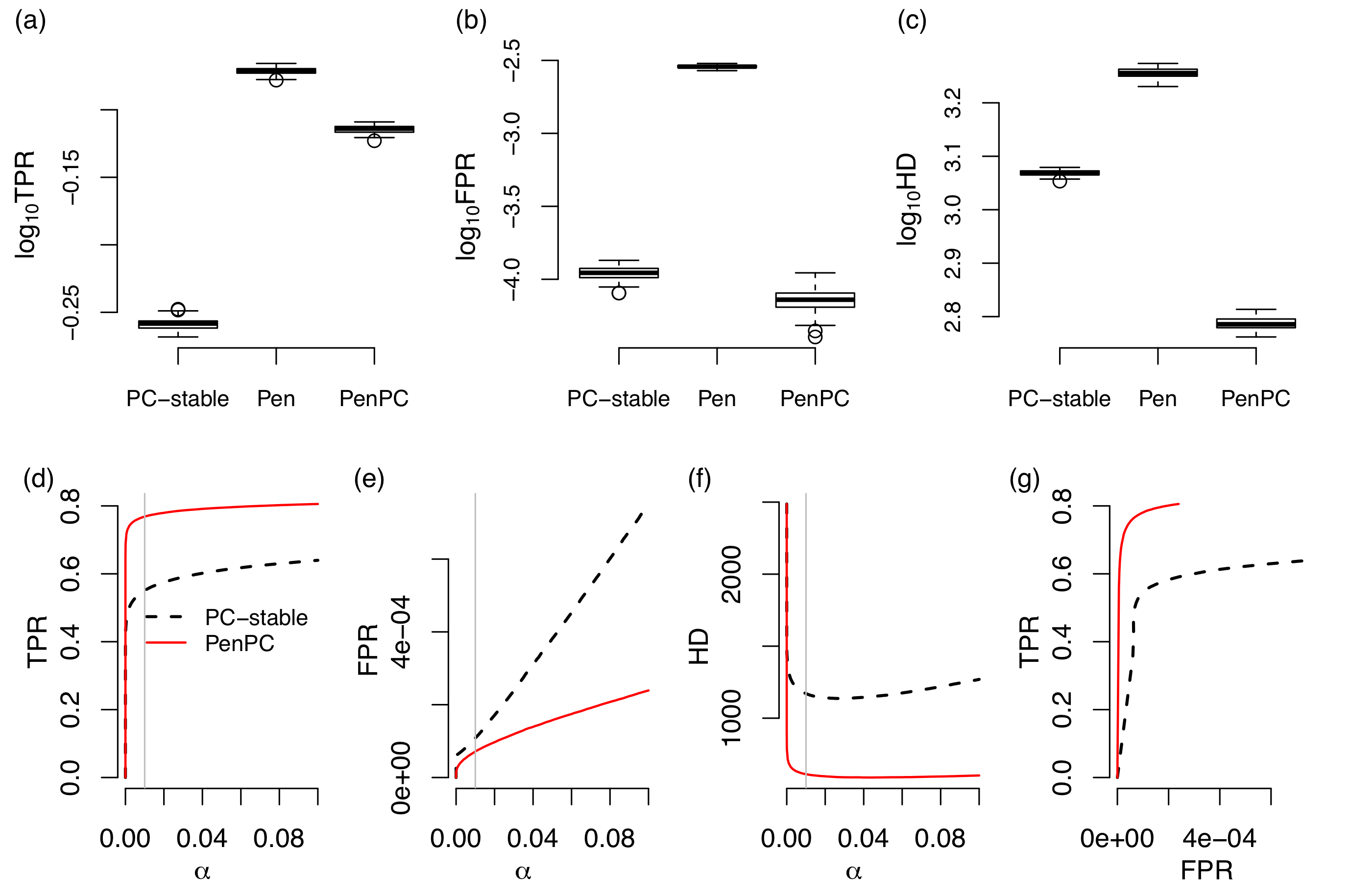}
\caption{Performance of ER model ($p=1000, n=300, p_E=0.005$). The
  upper panels are box plots (in $\log_{10}$ scale) of true positive
  rate (TPR) (a), false positive rate (FPR) (b) and hamming distance
  (HD) (c) from 100 replications at $\alpha=0.01$.  The lower panels
  are average true positive rate (d), false positive rate (e), and
  Hamming distance (f) from 100 replications when the tuning parameter
  $\alpha$ is changed from 0 to 0.1 (the grey vertical line are at
  $\alpha = 0.01$). ROC curves are shown in panel (g).}
\end{figure}

\begin{figure}[h!]
 \centering
\includegraphics[width=1\textwidth]{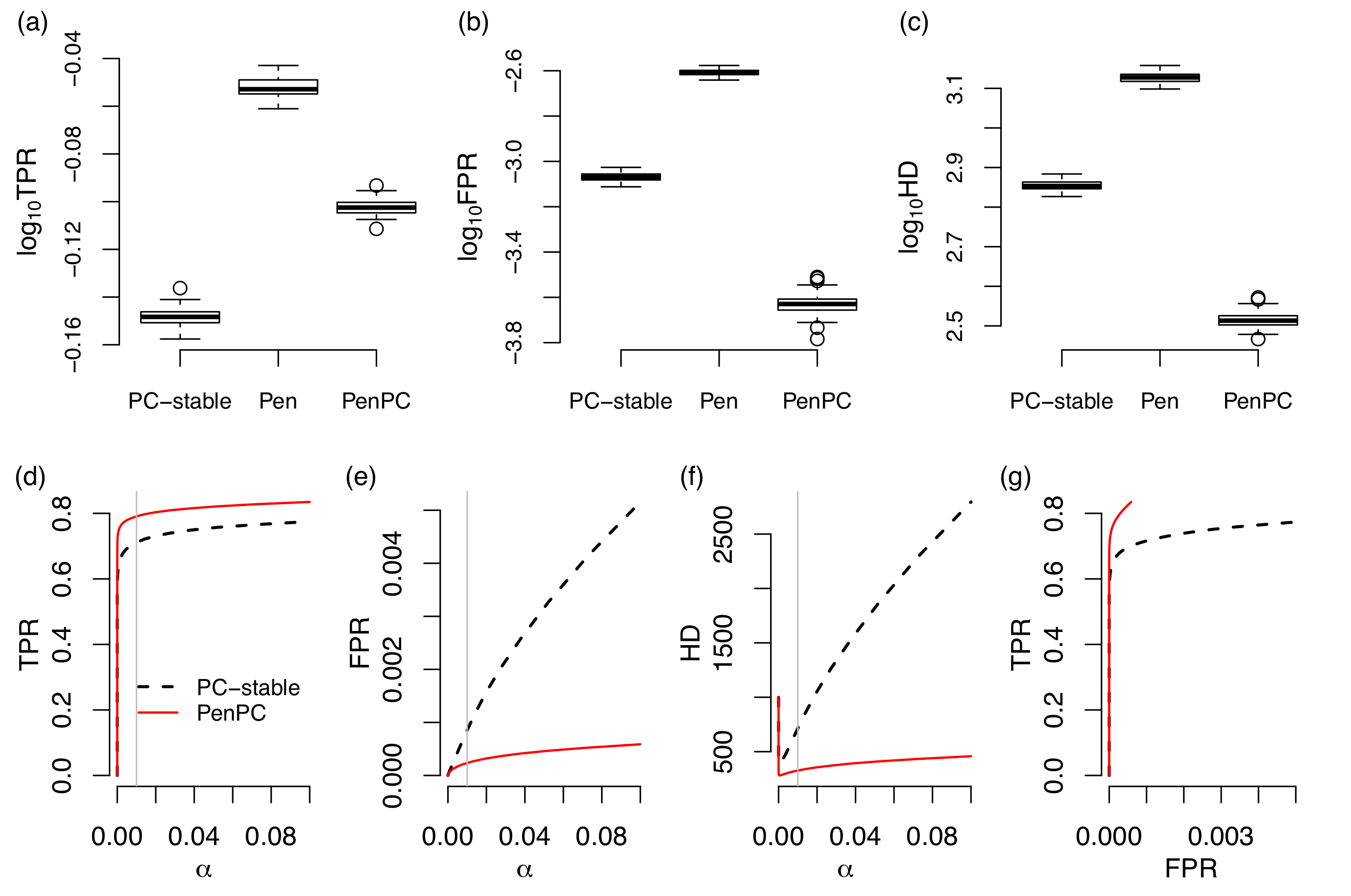}
\caption{Performance of BA model (p=1000,n=300,e=1). The upper panels
  are box plots (in $\log_{10}$ scale) of true positive rate (TPR)
  (a), false positive rate (FPR) (b) and hamming distance (HD) (c)
  from 100 replications at $\alpha=0.01$.  The lower panels are
  average true positive rate (d), false positive rate (e), and Hamming
  distance (f) from 100 replications when the tuning parameter
  $\alpha$ is changed from 0 to 0.1 (the grey vertical line are at
  $\alpha = 0.01$). ROC curves are shown in panel (g). }
\end{figure}

There are three tuning parameters. $\lambda$ and $\tau$ are tuning
parameters for the penalty function of the \texttt{PenPC} algorithm.
$\alpha$ is the p-value cutoff used by the PC-stable algorithm or our
modified PC-stable algorithm to declare conditional independence. We
chose $\lambda$ and $\tau$ by extended BIC \citep{chen2008extended},
and examined the results of PC or \texttt{PenPC} across various values
of $\alpha$. In the upper panels of Figure 3, we showed the
performances of three methods: PC (PC-stable algorithm), Pen
(penalized regression only, step 1 of the \texttt{PenPC}), and
\texttt{PenPC} when $\alpha=0.01$ and the skeleton was simulated by
the ER model. The penalized regression identifies more true positives
than the PC-stable algorithm, but also introduce more false positives
(Figure 3 (a-b)), while \texttt{PenPC} algorithm significantly reduces
the number of false positives, though some true positives are also
removed. At the end, the \texttt{PenPC} has the lowest number of false
positives plus false negatives, as measured by Hamming distance (HD)
(Figure 3 (c)). Figures 3(d-f) show that across various cutoff values
of $\alpha$, \texttt{PenPC} consistently has better performance than
the PC-stable algorithm. Finally, Figure 3(g) shows the ROC curves for
the \texttt{PenPC} and the PC-stable algorithms, which illustrate that
\texttt{PenPC} has better sensitivity and specificity than the
PC-stable algorithm regardless of the cutoff $\alpha$. Similar
conclusions can be drawn for the simulation results shown in Figure 4,
where the DAGs are simulated by the BA model.

\section{Application}
%
We applied the PC algorithm and the PenPC algorithm to study gene-gene
network using gene expression data from tumor tissue of breast cancer
patients. Gene expression were measured by RNA-seq
\citep{cancer2012comprehensive}. We quantified the expression of each
gene within each sample by log(total read count), or in short,
logTReC. We restricted our study on 550 female caucasian
samples. After removing genes with low expression across most samples,
we ended up with 18,827 genes. In this analysis, we focused on 410
genes from the cancer Gene Census in
\url{http://cancer.sanger.ac.uk/cancergenome/projects/census/}.  We
chose this relatively small gene set for two reasons. One is that it
is easier to exploit the results given that these genes have better
cancer-related annotations. The other reason is that we would like to
compare the results of the PC algorithm and the PenPC
algorithm. However, when we worked on a larger gene set of more than
8,000 genes, the PC algorithm took too much time to finish the
computation. We defer the discussion of computational efficiency in
the discussion section.

Several covariates may influence the correlations across genes. We
removed such effects by taking residuals of logTReC data for each gene
using a linear regression model with the following covariates: 75
percentile of logTReC per sample, which captures read depth, plate,
institution, age, and 6 genotype PCs.
\begin{figure}[h!]
  \centering
    \includegraphics[width=3.2in]{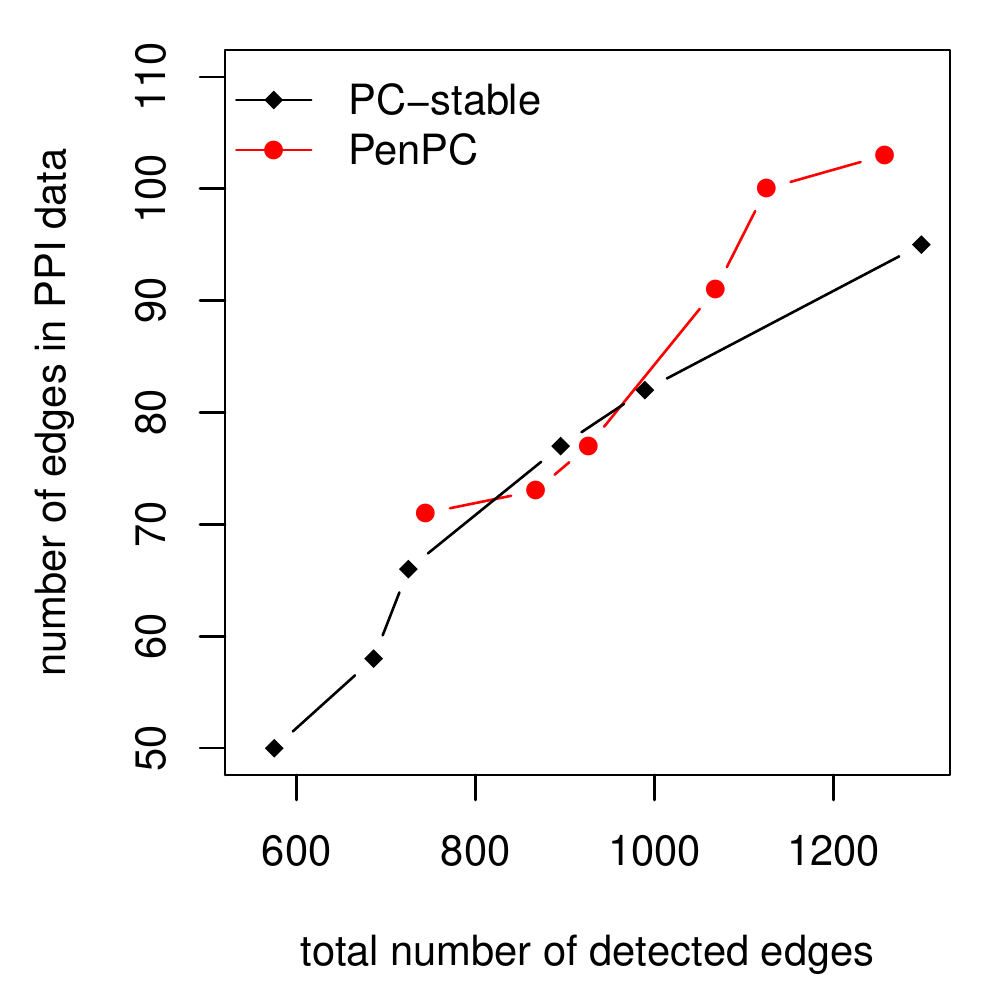}
      \caption{Comparing \texttt{PenPC} algorithm with PC-stable algorithm in terms of skeleton estimation by changing the significance levels for partial correlation testings, $\alpha = $0.0001, 0.0005, 0.001, 0.005, 0.01 and 0.05.}
      \label{fig:numsigPPI}
\end{figure}
Then for $\alpha =$ 0.0001, 0.0005, 0.001, 0.005, 0.01 and 0.05, we
estimated the skeleton by the PC-stable and \texttt{PenPC}
algorithms. The estimated skeletons were evaluated by comparing the
estimated edge sets with protein-protein interaction (PPI) database at
\url{http://www.pathwaycommons.org/pc2/downloads.html}, and we used
the protein annotations from the Universal Protein Resource
(\url{http://www.uniprot.org}). There were 3315 PPIs where both
proteins were matched to the 410 genes in our gene expression
data. Figure \ref{fig:numsigPPI} shows the total number of detected
edges versus the number of edges in PPI data. For both methods, the
total number of detected edges increase monotonically as $\alpha$
increases. The \texttt{PenPC} algorithm consistently detects more or
comparable number of edges than PC-stable algorithm, which reflects
the sensitivity, given the same total number of edges, which reflects
the specificity.

\section{Conclusions}
We propose a two-step approach, \texttt{PenPC} algorithm, to estimate
the skeletons of high dimensional DAGs. We show that the
\texttt{PenPC} algorithm provides asymptotically consistent estimate
of the skeleton of a high dimensional DAG. For fixed graphs, the
number of vertices $p_n$ could be exponential scale of the sample size
$n$. We also considered two commonly used random graph models and
discussed in detail the conditions under which the consistency
properties hold. The simulation studies and real data analysis show
that the network skeletons estimated by \texttt{PenPC} can be
substantially more accurate than those estimated by the PC-stable
algorithm. Although \texttt{PenPC} performs well for the scale-free
network, further improvement is possible by incorporating a
regularization method which prefers to the scale-free structure
\citep{liu2011learning} in the first step of the \texttt{PenPC}.

The acyclic assumption may appear restrictive for gene-gene network
since there may be feed back loops in gene expression regulation. One
solution is to use structure equation models (SEMs) where loops are
allowed \citep{li2006structural}.  Recently, a few methods have been
developed to add penalization into the SEM \citep{logsdon2010gene},
and we conjecture that adopting folded concave penalties in these
methods may further improve their performance. The other solution is
to construct Dynamic Bayesian Network using time course data
\citep{husmeier2003sensitivity}. This becomes a situation where the
natural ordering of the variables are available through time
information, and thus penalized regression itself is able to identify
the DAG skeleton through estimating conditional auto-regressive
correlations. The main challenge would be that the time course data
usually have limited number of time points and thus augmenting data
from other sources would be useful.

\begin{figure}[h!]
  \centering
  \includegraphics[width=3.2in]{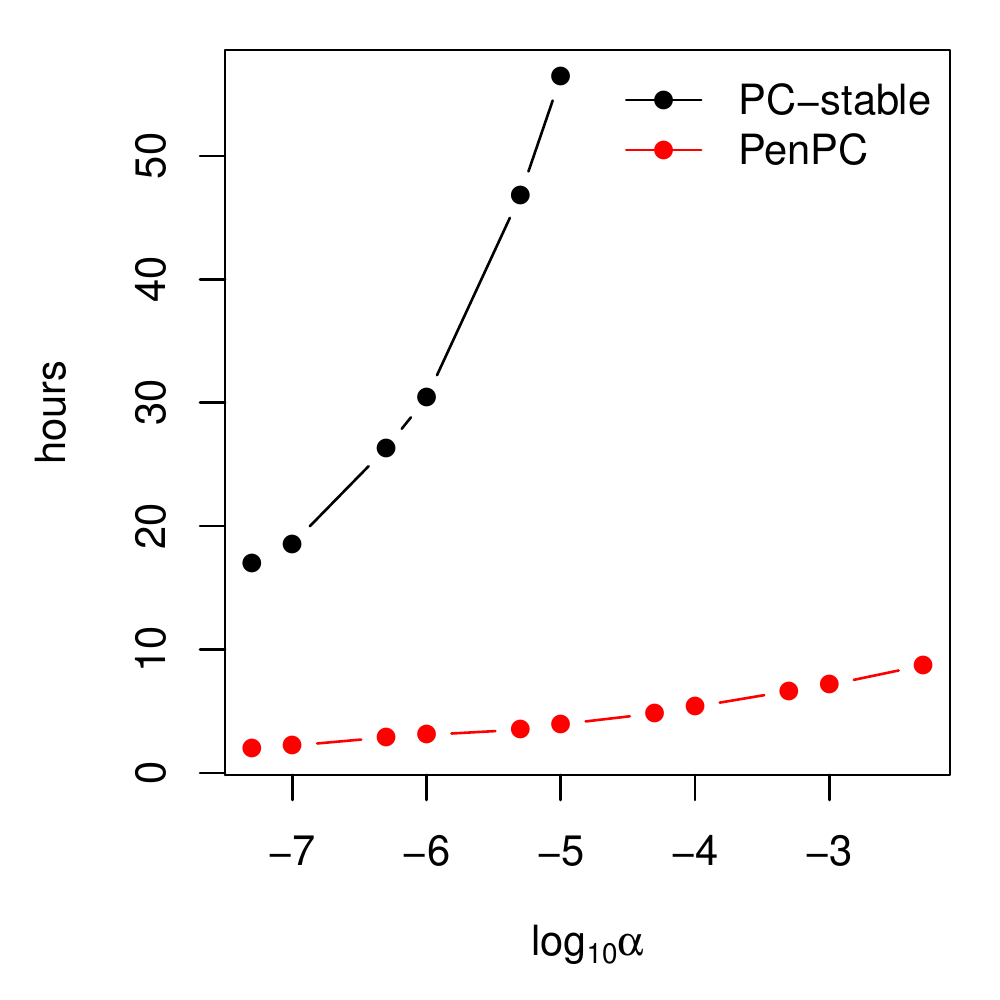}
      \caption{Computation time for PC-stable and step 2 of \texttt{PenPC}. }
      \label{fig:comptime}
\end{figure}

The computational efficiency of the PC-stable algorithm and our
modified PC-stable algorithm increases as the number of vertices
increases and as the p-value cutoff increases. When the dimension of
the problem becomes high enough, PC-stable algorithm becomes
computationally inefficient. We discuss the computational efficiencies
in two settings where $p=410$ or $p=8,261$. In our real data analysis
where $n=550$ and $p=410$. On average the step 1 of the \texttt{PenPC}
algorithm took 3 seconds for one penalized regression, including
searching for the best tuning parameter combination across a $100
(\lambda) \times 10 (\tau)$ two-dimensional grid. Thus the total
computational time is $3 \times 410 / 60 = 20.5$ minutes. As p-value
cutoff varies from $10^{-4}$ to $0.05$, the computational time of the
PC algorithm increases from 3 minutes to 54 minutes, while the
computational time of the 2nd step of the \texttt{PenPC} algorithm
increases from 17 seconds to 8 minutes. Overall the computational time
of the two methods are comparable and certainly \texttt{PenPC} is
computationally more attractive if one wants to examine the results
across multiple p-value cutoffs. We also examine the computational
efficiency when we expand the number of genes to $p$=8,261. The step 1
in \texttt{PenPC} algorithm took 128 seconds for one penalized
regression, including tuning parameter selection across a $100\times
10$ two-dimensional grid. This step, although computationally
expensive, can be easily paralleled. Given the GGM, the 2nd step of
the \texttt{PenPC} is computationally much more efficient than the
PC-stable algorithm (Figure~\ref{fig:comptime}). For example, with
p-value threshold varies from $10^{-7}$ to $10^{-5}$, the
computational time of the PC algorithm increases from 20 hours to 50
hours, and we did not run PC algorithm for p-value larger than
$10^{-5}$ due to high computational burden. In contrast, the
computation time of the \texttt{PenPC} remains below 10 hours even for
p-value cutoff $5 \times 10^{-3}$. All the computation are done in
Linux server with an 2.93 GHz Intel processor and 48GB RAM.

\FloatBarrier
\bibliographystyle{apalike}
\bibliography{skeleton}

\newpage

\appendix
\setcounter{section}{0}
\setcounter{equation}{0}
\setcounter{figure}{0}
\setcounter{table}{0}


 
\renewcommand{\thesection}{S.\arabic{section}}
\renewcommand{\thesubsection}{\thesection.\arabic{subsection}}
 
%
\makeatletter 
\def\tagform@#1{\maketag@@@{(S\ignorespaces#1\unskip\@@italiccorr)}}
\makeatother
 
\makeatletter
\makeatletter \renewcommand{\fnum@figure}
{\figurename~S\thefigure}
\makeatother
 
\renewcommand{\bibnumfmt}[1]{[S#1]}
\renewcommand{\citenumfont}[1]{\textit{S#1}}
 
\renewcommand{\figurename}{Figure}



\begin{center}
\textbf{\LARGE Supplementary Materials for ``\texttt{PenPC}: A Two-step Approach to Estimate the Skeletons of High Dimensional Directed Acyclic Graphs''}
\end{center}

\section{An example that neither covariance matrix nor concentration
matrix captures the network skeleton} \vspace{-0.3cm} Consider a
simple network of four nodes/variables $X$, $Y$, and $Z$ and $W$, with
the underlying network structure $X \rightarrow W \leftarrow Z
\leftarrow Y$, and we assume there is no any other (hidden)
variables. For illustration purpose, we assume the observations of
these four random variables are generated through the following
mechanism.
\begin{eqnarray} X = \epsilon_1, \ Y = \epsilon_2, \ Z &=& Y +
\epsilon_3, \ \textrm{and} \ W = X + Z + \epsilon_4 \label{eq:01}
\end{eqnarray} where $\epsilon_j$ are i.i.d. $N(0,1)$ for $1 \leq j
\leq 4$. Denote the covariance matrix and partial covariance matrix of
this system as $\bSigma$ and $\bOmega$, respectively. Note $\bOmega =
\bSigma^{-1}$, and $(i,j)$-th entry of $\bOmega$ indicates the
covariance of the $i$-th and the $j$-th variables, conditioning on all
the other covariates in this system. Let the connection matrix (i.e.,
skeleton) of this system be $\bXi$. Then we have:
\begin{displaymath} 
\small 
\bSigma = \bordermatrix{ & X & Y & Z & W \cr X & 1
    & 0 & 0 & 1 \cr Y & 0 & 1 & 1 & 1 \cr Z & 0 & 1 & 2 & 2 \cr W & 1
    & 1 & 2 & 4 \cr } \textrm{, } 
    \bOmega = \bordermatrix{ & X & Y & Z
    & W \cr X & 2 & 0 & 1 & -1 \cr Y & 0 & 2 & -1 & 0 \cr Z & 1 & -1 &
    2 & -1 \cr W & -1 & 0 & -1 & 1 \cr } \textrm{, }
     \bXi =
  \bordermatrix{ & X & Y & Z & W \cr X & 1 & 0 & 0 & 1 \cr Y & 0 & 1 &
    1 & 0 \cr Z & 0 & 1 & 1 & 1 \cr W & 1 & 0 & 1 & 0 \cr }.
\end{displaymath}\\
We see that neither $\bSigma$ nor $\bOmega$ gives us the correct
connection matrix of network structure $X \rightarrow W \leftarrow Z
\leftarrow Y$.

\section{The details of the \texttt{PenPC} algorithm}

In this section, we describe the step 2 of \texttt{PenPC}
algorithm. For any undirected graph $\G = (V,F_{\G})$, we define the following
quantities: 
\begin{itemize}
\item $\A_{\G, i,j} = \left[\adj(i,\G) \bigcup \adj(j,\G)\right]
  \setminus \{i,j\}$,
\item $\B_{\G,i,j} = \left[\adj(i,\G) \bigcap \adj(j,\G)\right]
  \setminus \{i,j\}$, and
  \item $\C_{\G,i,j} = \{k: k \in \A_{i,j} \ \bigcap \ (\B_{\G,i,j}\ \bigcup \ \Con_{\G}^{(i,j)}(\B_{\G,i,j})) \}$, where $\Con_{\G}^{(i,j)}(\B_{\G,i,j})$  is the set of vertices connected to any vertex in $\B_{\G,i,j}$ by a chain of any length from a subgraph of $\G$, which is created by  removing vertices $i$ and $j$ as well as any edges connected to $i$ or $j$.  
\end{itemize}

Then the algorithm is as follows. 

\begin{figure}[ht!]
\begin{boxedminipage}{\textwidth} \footnotesize
\noindent \textbf{Input:} GGM $\mC_{\mG} = (V, F_{\mG})$, which is obtained from the first step of the PenPC algorithm.\\ 
\textbf{Output:} Skeleton $\mG^u=(V,E^u)$ and separation set $S(i,j)$ for edges 
$i - j\notin E^u$ but $i - j \in F_{\mG}$.
        \begin{itemize} \vspace{-0.1cm}
         \item[1.] \textbf{Set} l = -1 and $\G = (V, F_{\G}) =\mC_{\mG}$, i.e., $F_{\G} = F_{\mG}$. 
	\vspace{-0.1cm}
         \item[2.] \textbf{For} any edge $i - j\in F_{\G}$,
                 \begin{itemize} \vspace{-0.3cm}
                 \item[2.1] If $X_i$ and $X_j$ are marginally independent, then \\
                                        - delete edge $i - j$ from $F_{\G}$, and\\ 
                                        - set $S(i,j) = \emptyset$.
                 \vspace{-0.3cm}
                 \end{itemize} \vspace{-0.1cm}
        \item[3.] \textbf{Repeat:} l = l+1 \vspace{-0.1cm}
                     \begin{itemize}
                     \item[3.1] $\tilde{\G}=\G$ \vspace{-0.1cm}
                     \item[3.2] \textbf{For} any edge $i - j\in F$ such that $|\C_{\tilde{\G}, i,j}|\geq l$. 
                     \vspace{-0.1cm}
                           \begin{itemize}
                           \item[3.2.1] \textbf{Repeat:} Select 
                           $\Gamma\subseteq\C_{\tilde{\G}, i,j}$ with $|\Gamma|=l$.
                                      \begin{itemize}
                                       \item[3.2.1.1] Set $\Kr=\A_{\tilde{\G}, i,j} \backslash \Gamma$.
                                        \item[3.2.1.2] If $X_i$ and $X_j$ are conditionally independent given $\{X_k:k\in \Kr\}$, then\\ 
                                        - delete edge $i - j$ from $F_{\G}$, and\\ 
                                        - $S(i,j) = \Kr$.
                            \end{itemize}
                           \item[3.2.2] \textbf{Until:} The edge $i - j$ is deleted or all $\Gamma$ with $|\Gamma|=l$ have been examined. 
                           \end{itemize} \vspace{-0.1cm}
                     \end{itemize} \vspace{-0.3cm}
        \item[4.] \textbf{Until:} for each $i - j\in F_{\G}$, $|\C_{\G, i,j}|<l$. 
        \item[5.] \textbf{Set} $\mG^u =(V,E^u) = \G$, i.e., $E^u = F_{\G}$. 
        \end{itemize} \normalsize
\end{boxedminipage}
\caption{The second step of the \texttt{PenPC} algorithm. In steps 3.1-3.2, we save the current graph $\G$ to $\tilde{\G}$, and all the conditional independence tests are based on $\tilde{\G}$ while $\G$ is being updated. This is the ``stable'' part of the PC-stable algorithm, so that the order of the edges being tested does not matter.}
\label{Fig:orderindepPenPC}
\end{figure} \clearpage

\section{The deterministic rules to extend a skeleton to a CPDAG} 
These deterministic rules have been described in \cite{kalisch2007estimating} and
\cite{pearl2009causality}. Given the skeleton $\mG^u$ and the
separation sets $S(i,j)$ for all missing edges between nodes $i$ and
$j$, the arrow orientation of the skeleton proceeds in two step: (1)
determination of the $v$-structure and (2) completion of the partially
directed graph (PDAG).
\begin{itemize}
\item[step 1] For each pair of nonadjacent vertices $i$ and $j$ with
common neighbor $k$, add arrow heads pointing at $k$, $i\rightarrow k
\leftarrow j$ if $k\notin S(i,j)$.
\item[step 2] In the PDAG from step 1, following four rules are
repeatedly applied to obtain maximally oriented pattern.
\begin{itemize}
\item[rule 1:] Orient $j-k$ into $j\rightarrow k$ whenever there is an
arrow $i\rightarrow j$ such that $i$ and $k$ are nonadjacent.
\item[rule 2:] Orient $i-j$ into $i\rightarrow j$ whenever there is a
chain $i\rightarrow k \rightarrow j$.
\item[rule 3:] Orient $i-j$ into $i\rightarrow j$ whenever there are
two chains $i-k_1\rightarrow j$ and $i-k_2\rightarrow j$ such that
$k_1$ and $k_2$ are nonadjacent.
\end{itemize}
\end{itemize} The repeated application of these rules results in
orienting all arrows that are common for all the DAGs within the same Markov equivalent class.

\clearpage
\section{Supplementary Figures}

\begin{figure}[!ht]
\caption{Histograms of the degree $\nu$ under BA model with $p=1000$
and $e=2$ and the $\log_{10}$ scale density of $\log_{10}\nu$ in the
subplot.}  \centering
\includegraphics[width=0.5\textwidth]{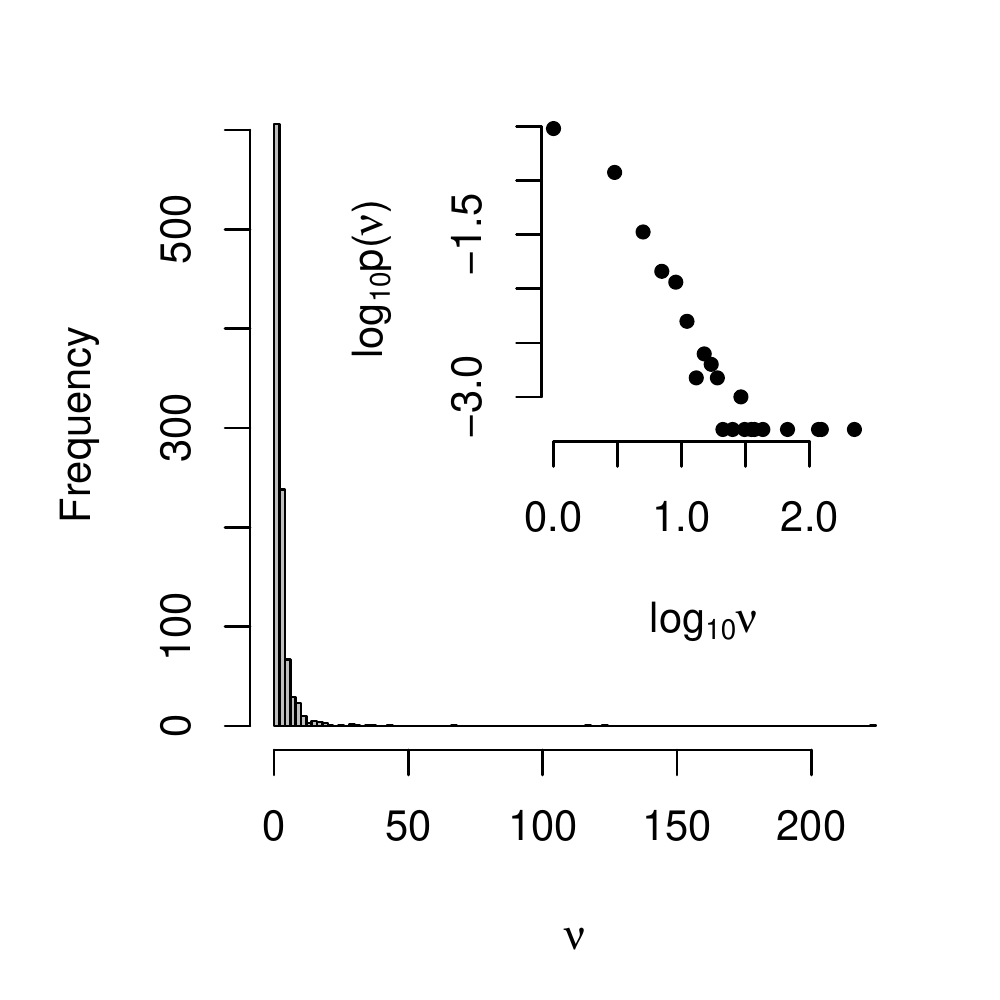}
\end{figure}

\begin{figure}[!ht]
\caption{ER model ($p=11, n=100, p_E=0.2$)} \centering
\includegraphics[width=1\textwidth]{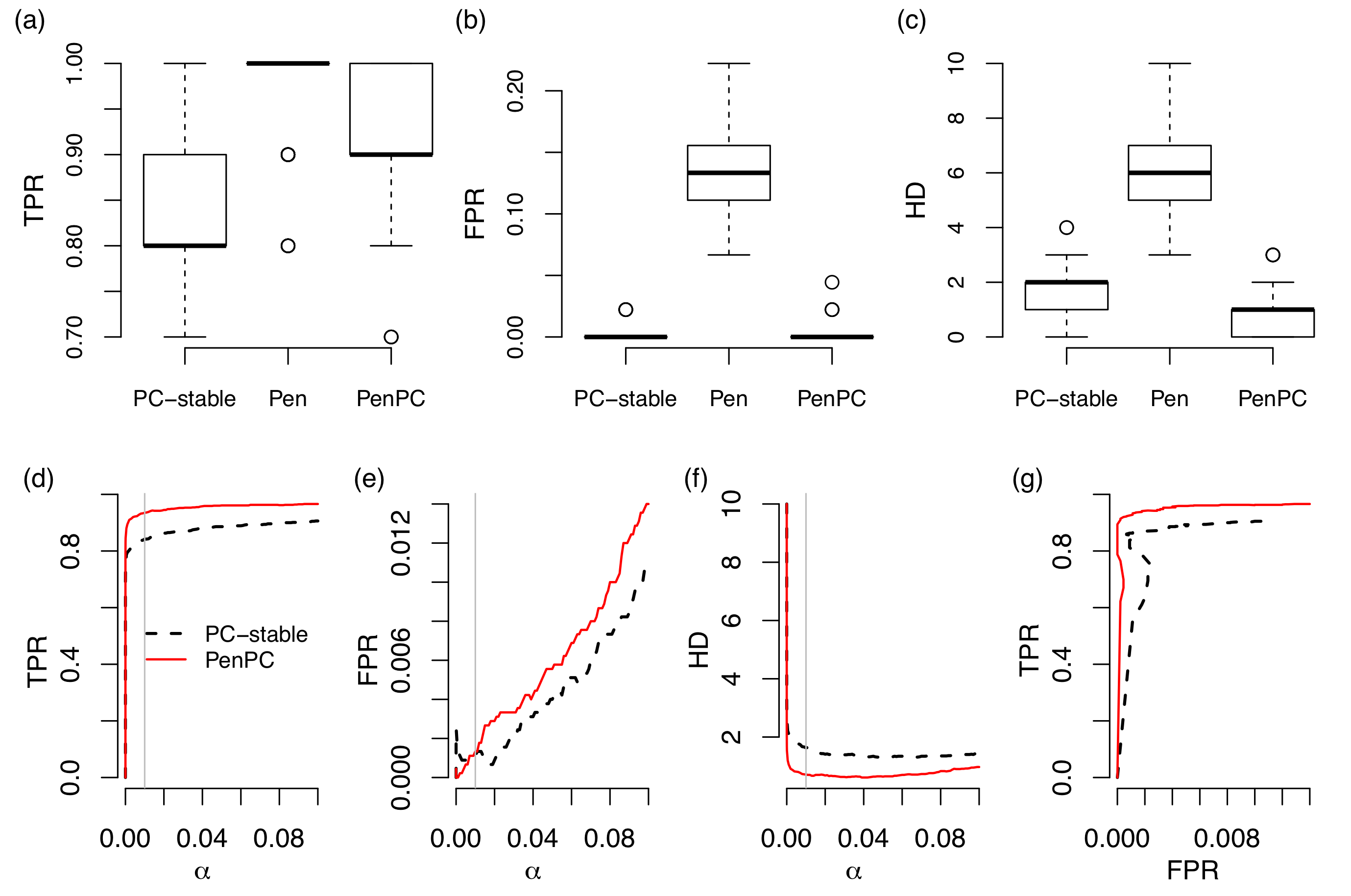}
\end{figure}

\begin{figure}[!ht]
\caption{ER model ($p=100, n=30, p_E=0.02$)} \centering
\includegraphics[width=1\textwidth]{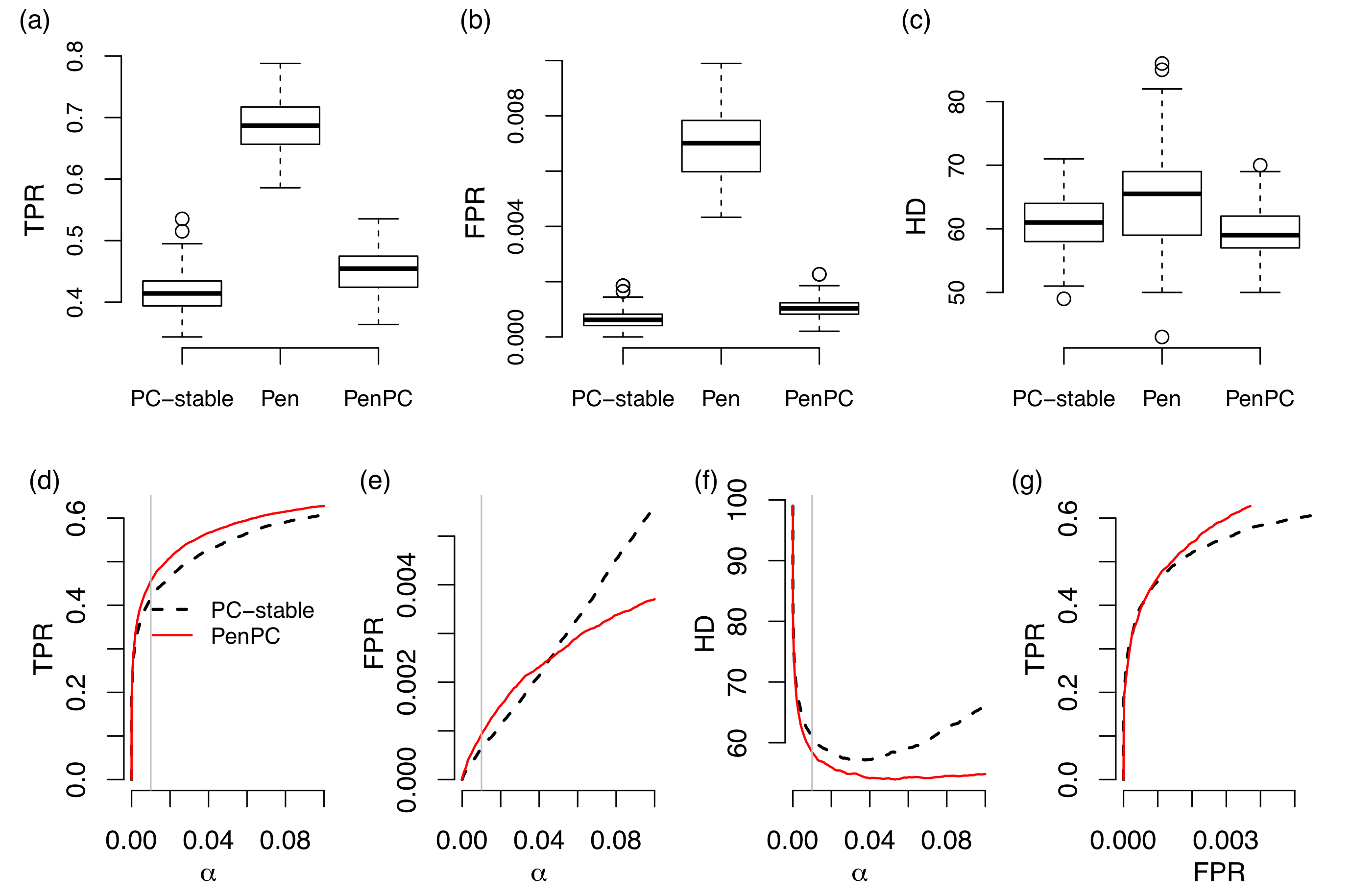}
\end{figure}

\begin{figure}[!ht]
\caption{ER model ($p=100, n=30, p_E=0.03$)} \centering
\includegraphics[width=1\textwidth]{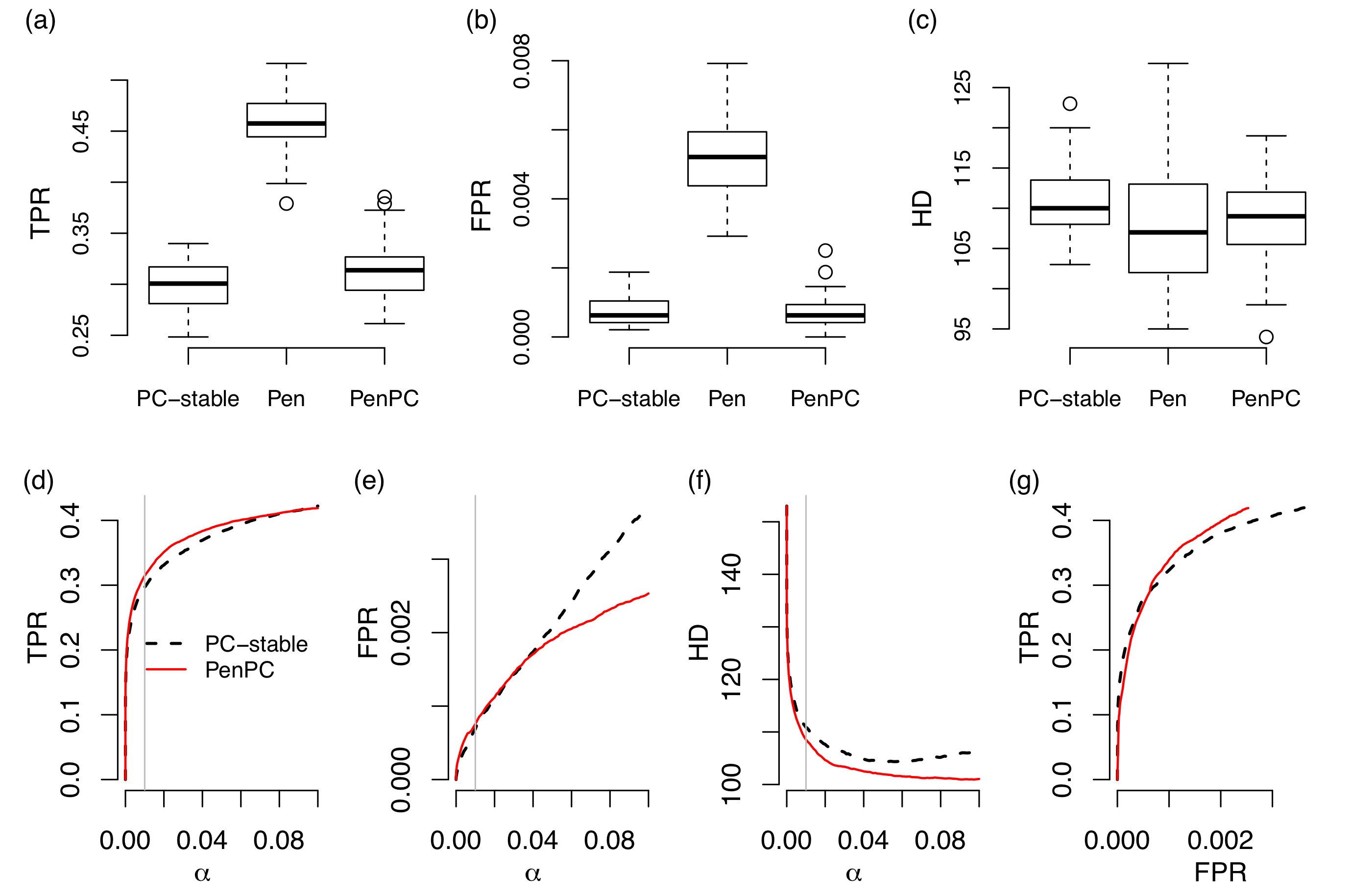}
\end{figure}

\begin{figure}[!ht]
\caption{ER model ($p=100, n=30, p_E=0.04$)} \centering
\includegraphics[width=1\textwidth]{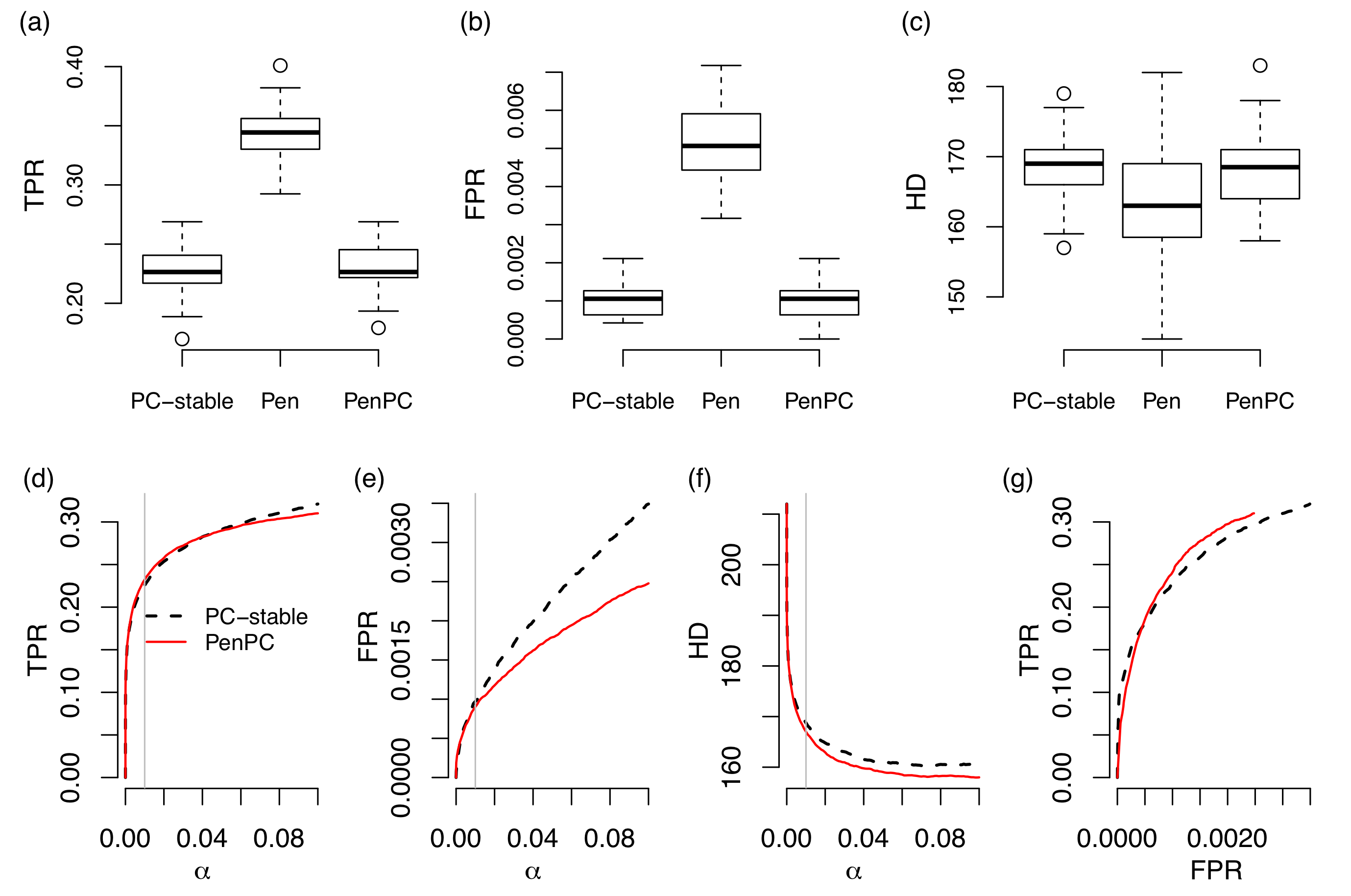}
\end{figure}

\begin{figure}[!ht]
\caption{ER model ($p=100, n=30, p_E=0.05$)} \centering
\includegraphics[width=1\textwidth]{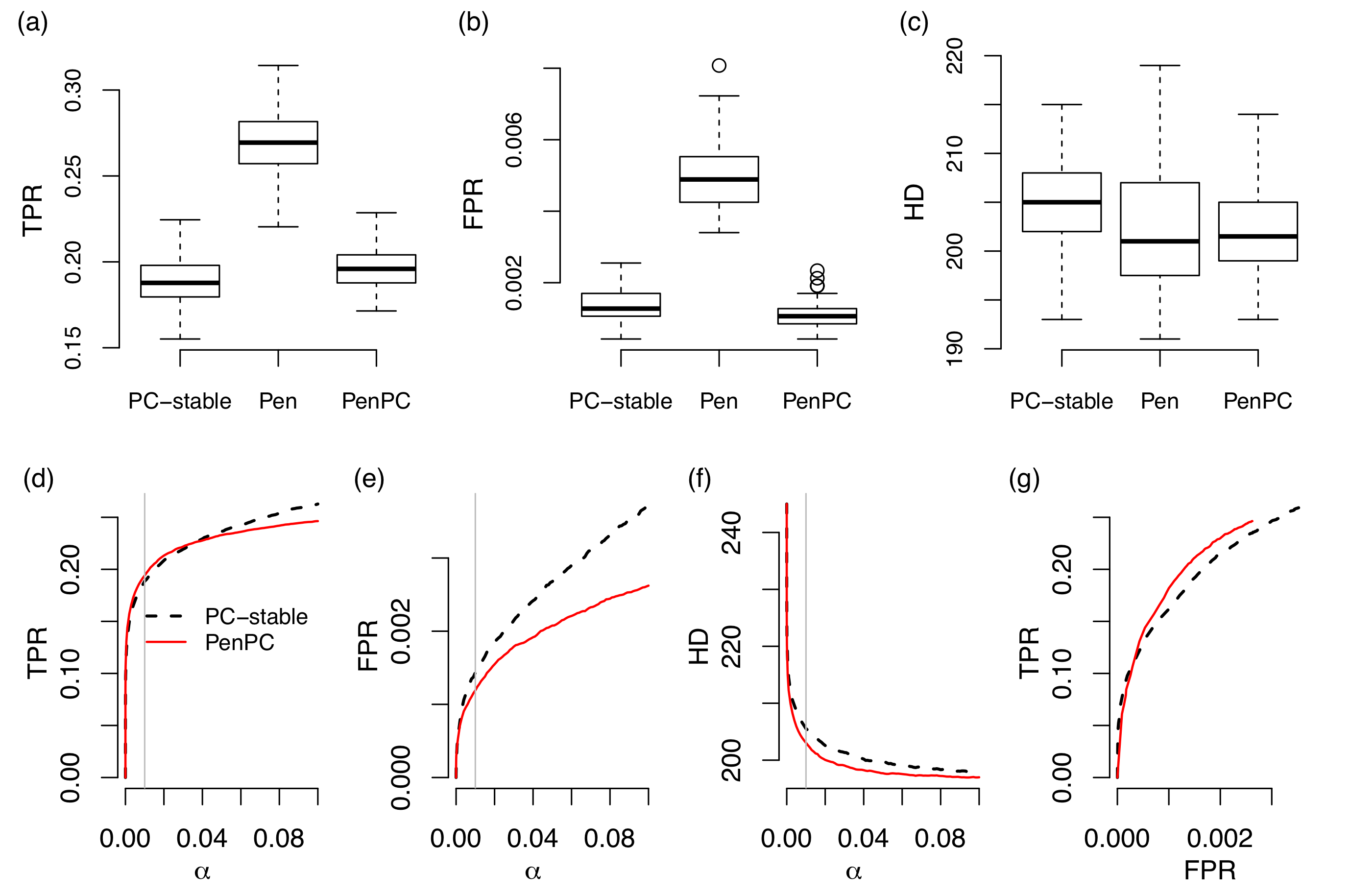}
\end{figure}

\begin{figure}[!ht]
\caption{ER model ($p=1000, n=300, p_E=0.002$)} \centering
\includegraphics[width=1\textwidth]{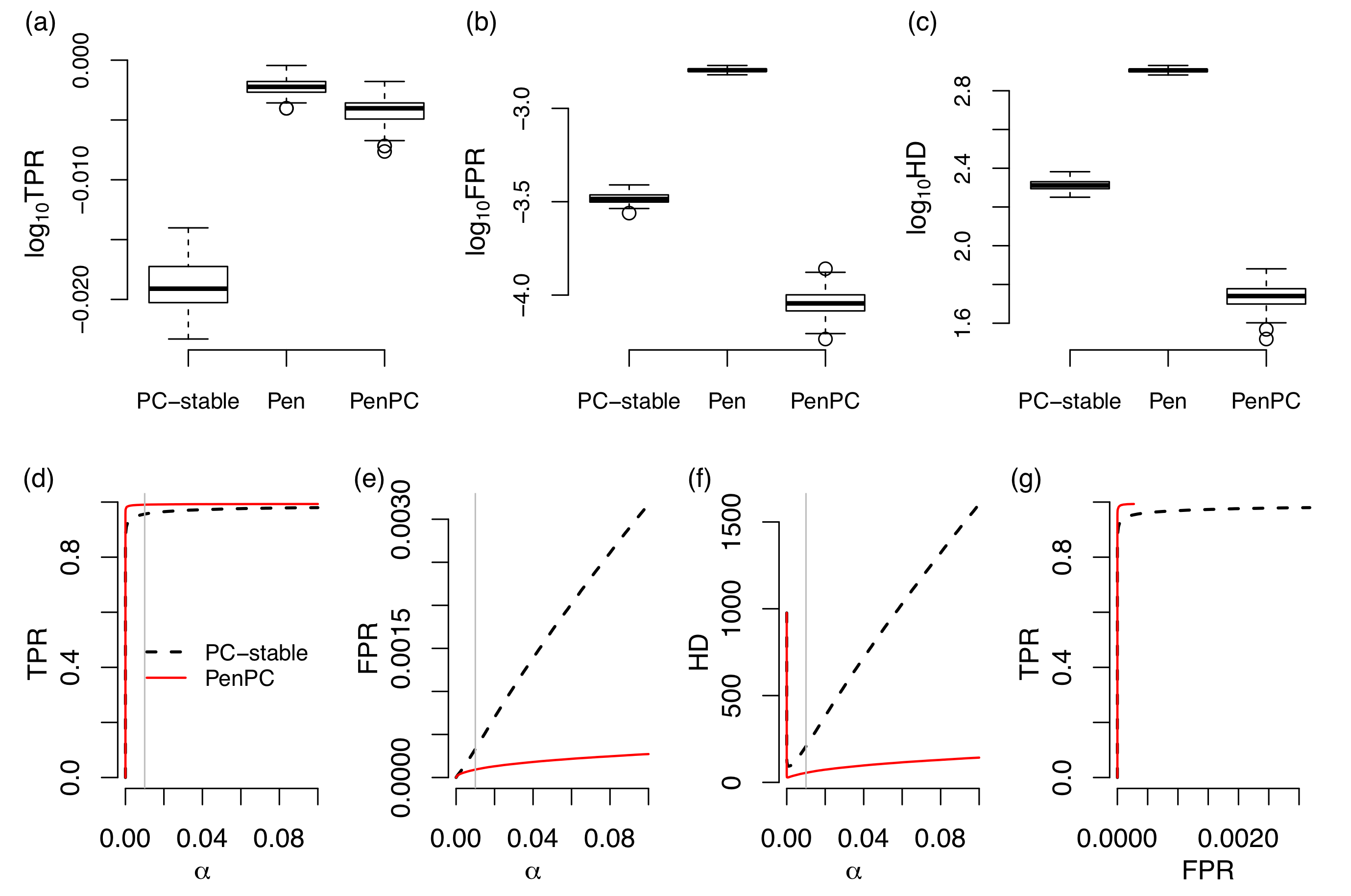}
\end{figure}

\begin{figure}[!ht]
\caption{ER model ($p=1000, n=300, p_E=0.01$)} \centering
\includegraphics[width=1\textwidth]{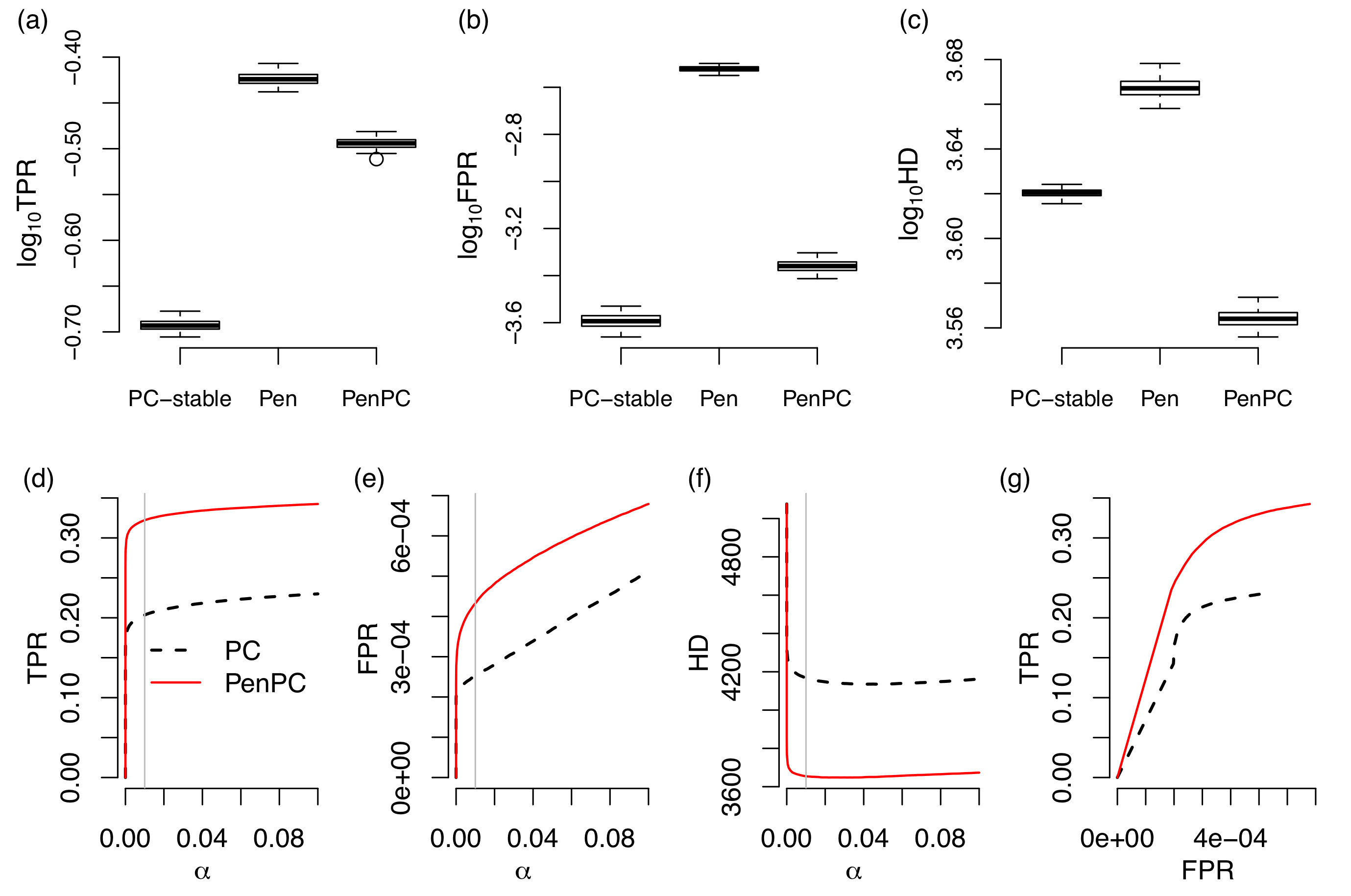}
\end{figure}

\begin{figure}[!ht]
\caption{BA model ($p=11, n=100, e=1$)} \centering
\includegraphics[width=1\textwidth]{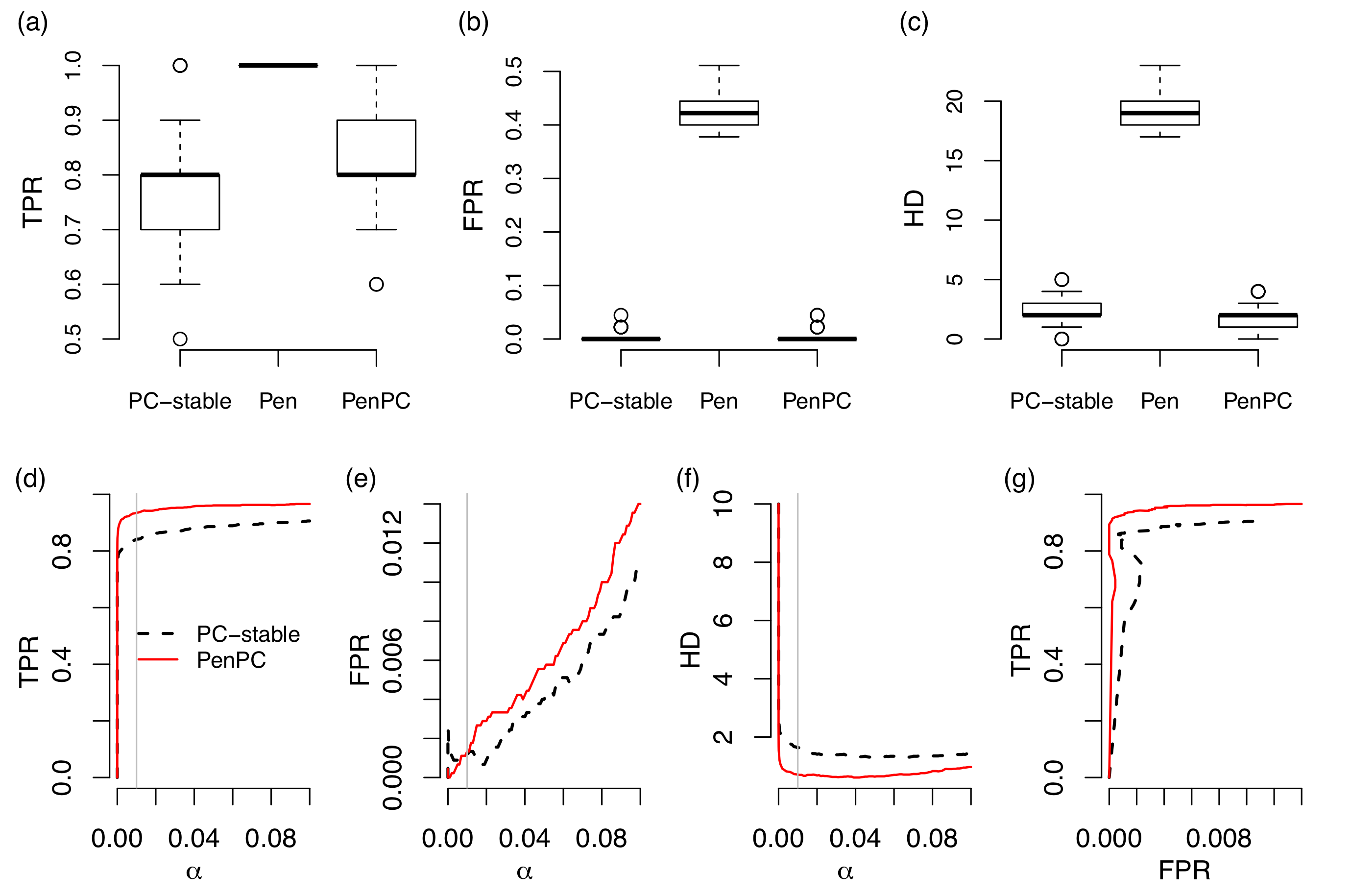}
\end{figure}

\begin{figure}[!ht]
\caption{BA model ($p=11, n=100, e=2$)} \centering
\includegraphics[width=1\textwidth]{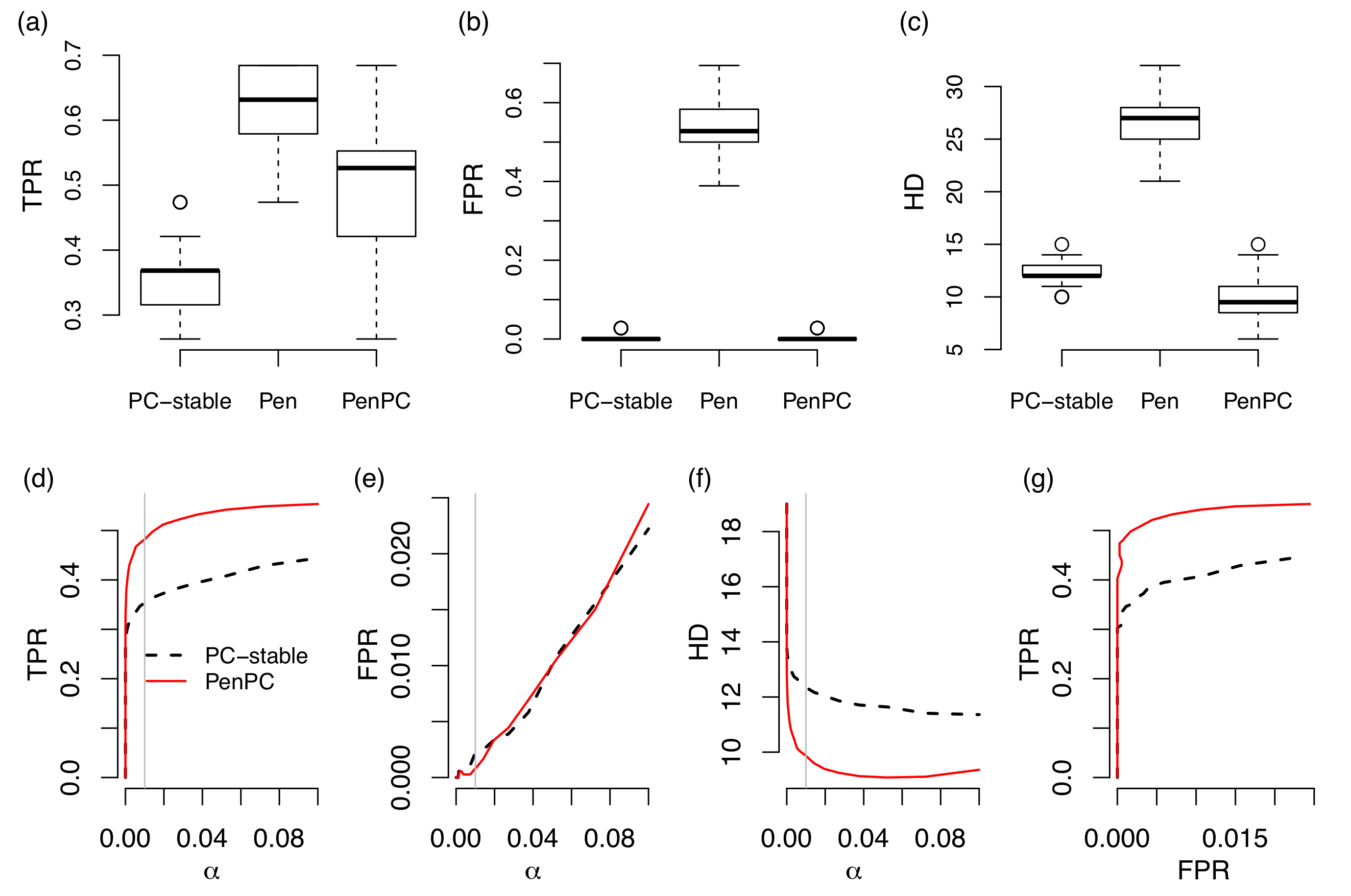}
\end{figure}

\begin{figure}[!ht]
\caption{BA model ($p=100, n=30, e=1$)} \centering
\includegraphics[width=1\textwidth]{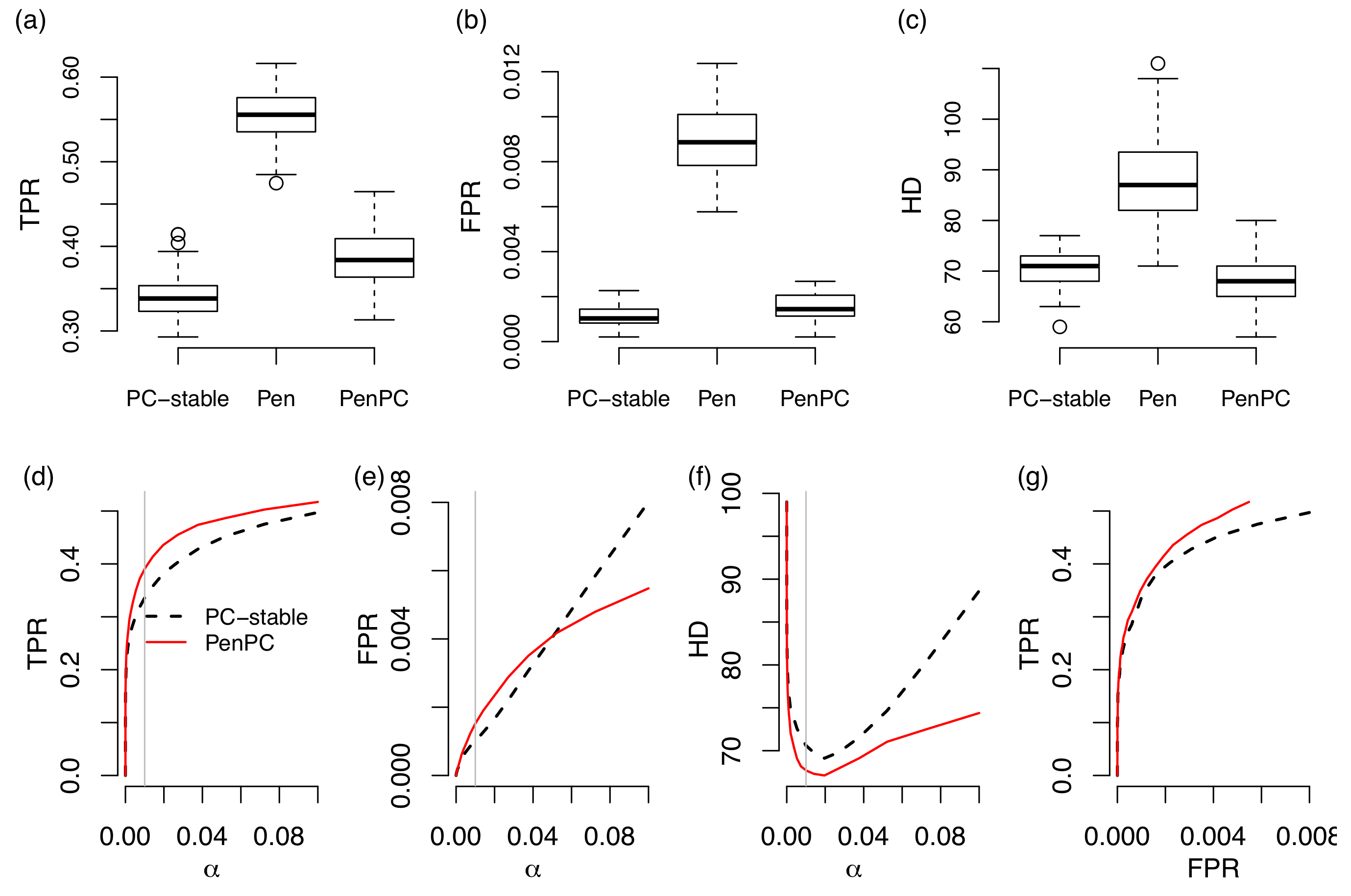}
\end{figure}

\begin{figure}[!ht]
\caption{BA model ($p=100, n=30, e=2$)} \centering
\includegraphics[width=1\textwidth]{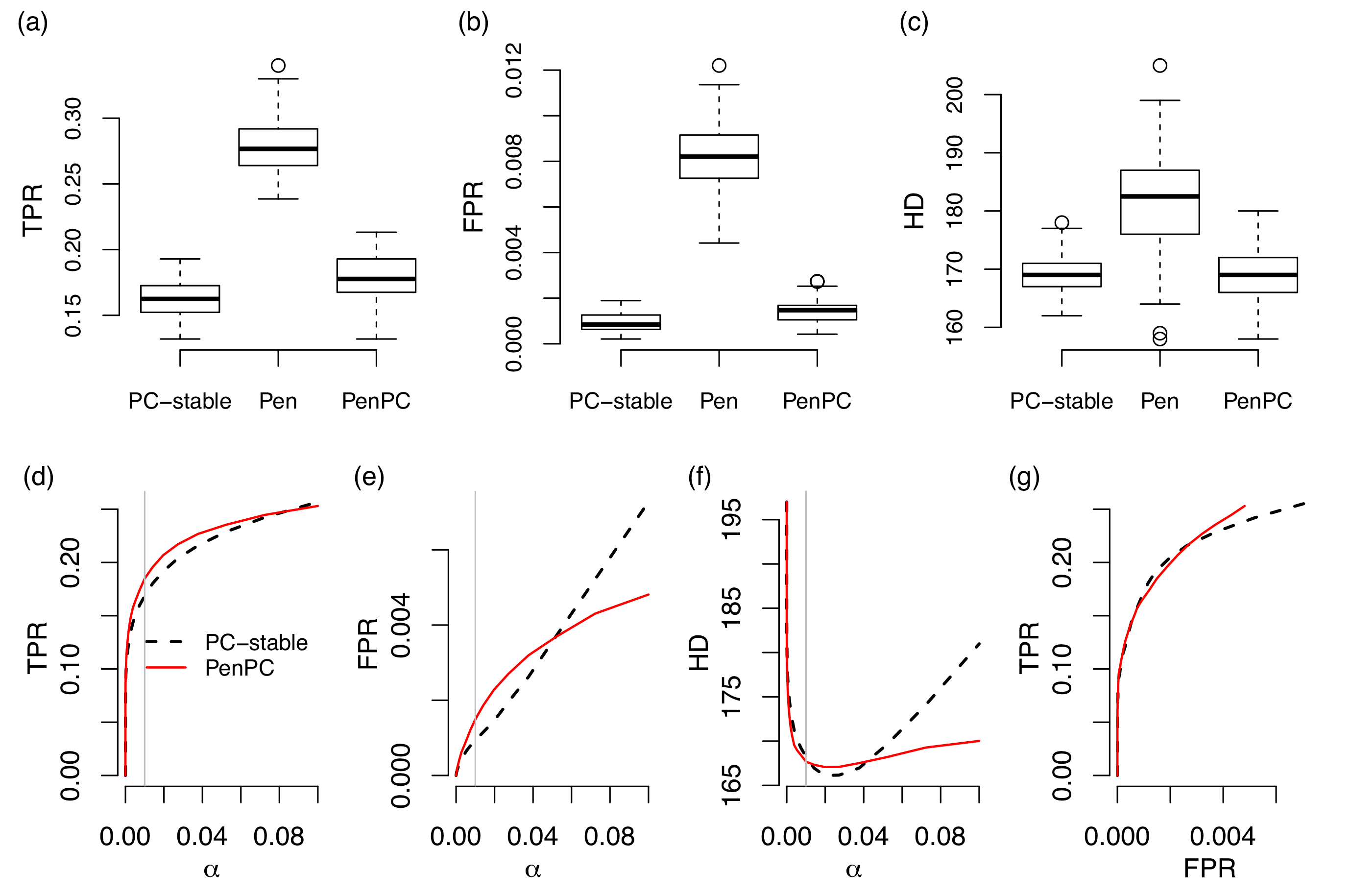}
\end{figure}

\begin{figure}[!ht]
\caption{BA model ($p=1000, n=300, e=1$)} \centering
\includegraphics[width=1\textwidth]{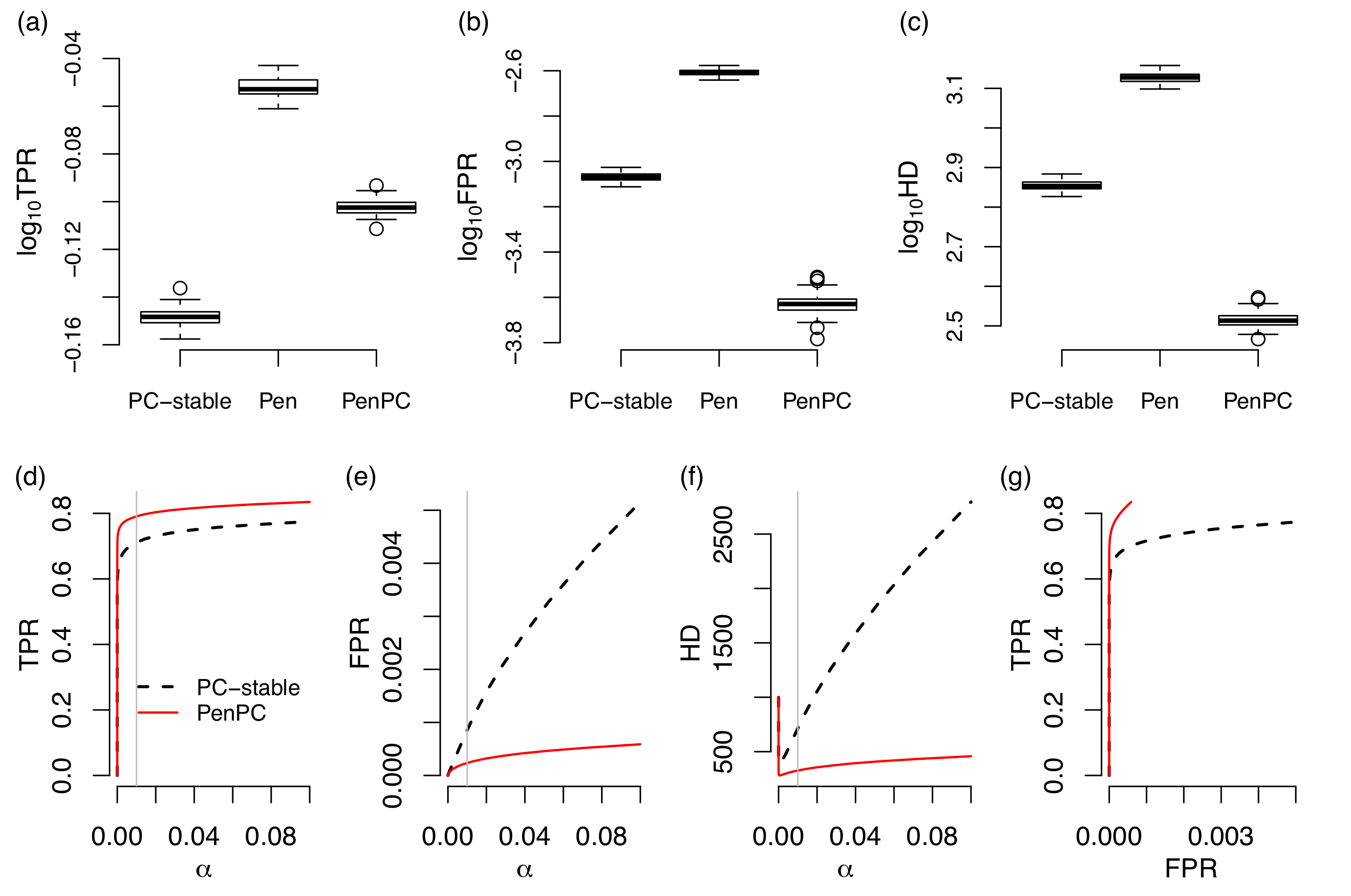}
\end{figure}

\begin{figure}[!ht]
\caption{BA model ($p=1000, n=300, e=2$)} \centering
\includegraphics[width=1\textwidth]{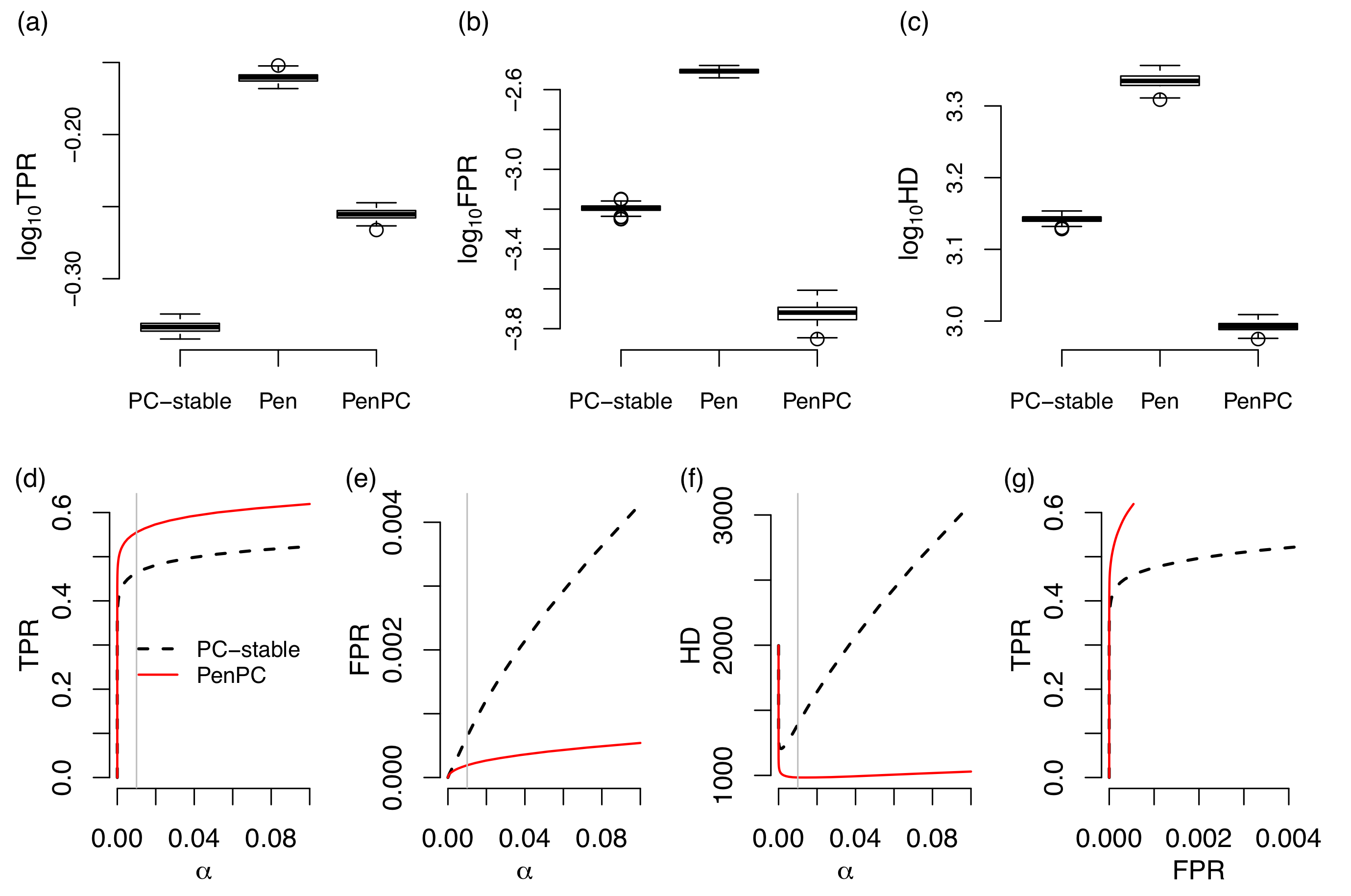}
\end{figure}

\clearpage


\section{Proofs}


\subsection{Lemma~5}

The following lemma is needed for proof of Theorem 1. It provides a
sufficient condition for strict local minimizer $\hat{\vb}_i$ of
equation (3) in the main text.\\

\noindent Lemma 5: Assume that $p(t;\lambda,\tau) = \lambda
\rho(t;\tau)$ satisfies Condition 1. Define $\bar{\rho}(t;\tau) =
\sgn(t)\rho'(\lvert t\rvert;\tau)$, $t\in \RR$ and
$\bar{\rho}(\vt;\tau) = (\bar{\rho}(t_1;\tau), \ldots,
\bar{\rho}(t_q;\tau))^\trans$, $\vt=(t_1,\ldots,t_q)^\trans$.  Then
$\hat{\vb}_i\in \RR^{p_n-1}$ is a strict local minimizer of
\[Q(\vb_i) = \frac{1}{2}(\vx_i -
\vX_{-i}\vb_i)^\trans(\vx_i-\vX_{-i}\vb_i) + n\sum_{j \neq i}
p(|b_{i,j}|;\lambda_i,\tau_i)\] if
  \begin{align} & \vXcr_{i1}^\trans (\vx_i - \vX_{-i}\hat{\vb}_i) =
n\lambda_i\bar{\rho}(\hat{\vb}_{i1};\tau_i) , \label{eq:1} \\
  & \lVert
\vXcr_{i2}^\trans(\vx_i-\vX_{-i}\hat{\vb}_i)\rVert_\infty <
p'(0+; \lambda_i, \tau_i), \label{eq:2}\\ &\lVert
(\vXcr_{i1}^\trans\vXcr_{i1})^{-1}\rVert_2 < 1/(n \kappa(\hat{\vb}_{i1}; \lambda_i, \tau_i)), 
\label{eq:3}
\end{align} 
where $\kappa(\bv; \lambda_i, \tau_i) = \lim_{\epsilon\rightarrow 0 +}  
	\max_{1\leq j\leq r} \sup_{t_1<t_2\in (\lvert v_j\rvert - \epsilon,\lvert v_j \rvert +
     \epsilon )} -\frac{p'(t_2; \lambda_i, \tau_i) -
     p'(t_1;\lambda_i, \tau_i)}{t_2-t_1}$ for any vector $\bv=(v_1,...,v_r)^\trans\in
  \mathbb{R}^{r}$, $\hat{\vb}_{i1}$ is the subvector of $\hat{\vb}_i$'s nonzero components. 
On the other hand, if $\hat{\vb}_i$ is a
local maximizer of $Q(\vb_i)$, then it must satisfy
(\ref{eq:1})-(\ref{eq:3}) with strict inequalities replaced by non-strict
inequalities.

Lemma 5 is a special case of the Theorem 1 in \cite{fan2011nonconcave} and thus we skip the proof.

\subsection{Proof of Theorem~1}

  For any fixed $i\in V_n$, $\vx_i$ is a $n\times 1$ response vector
and $\vX_{-i}$ is a $n\times q$ covariate matrix with $q=p_n-1$
corresponding to vertices $V_n\setminus \{i\}$. Let $\Sr_i =
\textrm{supp}(\vb_i)$ to be the support of the true regression
coefficient $\vb_i$ with $|\Sr_i|=s_i$. Define $\bxi_i = (\xi_{i1},
..., \xi_{iq})^\trans = \vX_{-i}^\trans(\vx_i -
\vX_{-i}\vb_i)=\vX_{-i}^\trans \beps_i$ where $\beps_i \sim
N_n(0,\sigma_i^2I_n)$ for $n\times n$ identity matrix $I_n$. Let
$\bxi_{i1}$ and $\bxi_{i2}$ to be the non-joint sub-vectors with
indices partitioned by $\Sr_i$. Define the event
\begin{equation}\label{eqn:eps} \Er_i = \left\{ \|\bxi_{i} \|_\infty
\leq \sigma_i n^{1/2 + a/2} \sqrt{\log(n)} \right\}.
\end{equation} We first consider the property of penalized regression
on $\Er_i$. Lemma~5 gives sufficient conditions of a local
minimizer. We prove that within the hypercube
\begin{equation}\label{eq:nr} \Nr_i = \{ \bbeta=(\bbeta_1^
\trans,\bbeta_2^\trans)^\trans\in\RR^q: \|\bbeta_{1} - \vb_{i1}
\|_\infty \leq Cn^{-d_2},\ \bbeta_{2} = 0\},
\end{equation} there is a solution $\hat{\vb}_i$ that satisfy
(\ref{eq:1}) and (\ref{eq:2}), and equation (\ref{eq:3}) of
Lemma~5 holds by Assumption (A5).

\noindent{\underline{Step 1: Find a solution to (\ref{eq:1}) in
$\Nr_i$.}}

We will prove that conditioning on $\Er_i$, there is a
solution $\hat{\vb}_{i1}\in \mathbb{R}^{s_i}$ for equation
(\ref{eq:1}) of Lemma~5 which is equivalent to
\[\hat{\vb}_{i1} = \vb_{i1} + (\vXcr_{i1}^\trans\vXcr_{i1})^{-1} \{
\vXcr_{i1}^\trans\beps -
n\lambda_i\bar{\rho}(\hat{\vb}_{i1};\tau_i)\}.\] Suppose that $\bbeta
= (\bbeta_{1}^\trans,\bbeta_{2}^\trans)^\trans \in \mathbb{R}^q$ has
the same partition as $\vb_i =
(\vb_{i1}^\trans,\vb_{i2}^\trans)^\trans$. Let $\vu_i =
(\vXcr_{i1}^\trans\vXcr_{i1})^{-1}[\vXcr_{i1}^\trans\beps -
n\lambda_i\bar{\rho}(\bbeta_1;\tau_i)]$, and $\phi(\bbeta_1) =
\bbeta_1 - \vb_{i1} - \vu_i$, where
$\bbeta_1=(\beta_{1,1},\ldots,\beta_{1,s_i})^\trans \in \RR^{s_i}$ and
$\vb_{i1}=(b_{i1,1},\ldots,b_{i1,s_i})\in \RR^{s_i}$. It suffies to
show that there is a solution to $\phi(\bbeta_1)=\vzero$ in
$\Nr_i$. Suppose $\|\vu_i\|_{\infty} = o(n^{-d_2})$. For sufficiently
large $n$, if $\beta_{1,j} - b_{i1,j} = Cn^{-d_2}$,
$\phi_j(\bbeta_1)\geq Cn^{-d_2} - \| \vu_i\|_{\infty} > 0$.  If
$\beta_{1,j} - b_{i1,j} = -Cn^{-d_2}$, $\phi_j(\bbeta_1)\leq
-Cn^{-d_2} + \|\vu_i\|_{\infty} < 0$.  By the continuity of function
$\phi(\bbeta_1)$ and Miranda's existence theorem, there is a solution
for $\phi(\bbeta_1)=\vzero$ in $\Nr_i$.

Now we prove $\|\vu_i\|_\infty=o(n^{-d_2})$. For any
$\bbeta=(\bbeta_1^\trans,\bbeta_2^\trans)^\trans\in \Nr_i$, $\lvert
\beta_{1,j}\rvert \geq \lvert b_{i1,j} \rvert - \delta_n$ where
$\delta_n$ is defined in Assumption (A3), and thus
\[\min_{j=1,\ldots,s_i} \lvert \beta_{1,j} \rvert \geq
\min_{j=1,\ldots,s_i} \lvert b_{i1,j} \rvert - \delta_n \geq
\delta_n.\] By monotonicity of $\rho(t;\tau)$ in Condition 1, $\|
\bar{\rho}(\bbeta_1;\tau_i)\|_\infty \leq
\rho'(\delta_n;\tau_i)$. Therefore, on $\Er_i$,
\[\| \vXcr_{i1}^\trans\beps - n\lambda_i\bar{\rho}(\bbeta_1;\tau_i) \|_\infty \leq
\sigma_i n^{1/2+a/2}\sqrt{\log(n)} + np'(\delta_n;\lambda_i,\tau_i).\] 
Then by Assumption (A5),
\[\| \vu_i \|_\infty \leq \sigma_i n^{-1/2+a/2+s_0} \sqrt{\log(n)} +
n^{s_0}p'(\delta_n;\lambda_i,\tau_i).\] By Assumption (A3),
$\sigma_in^{-1/2+a/2+s_0}\sqrt{\log(n}) = o(n^{-d_2})$ and by
Assumption (A4), $n^{s_0}p'(\delta_n;\lambda_i,\tau_i) =
o(n^{-d_2})$. Therefore, $\| \vu_i \|_\infty = o(n^{-d_2})$. 

\noindent\underline{Step 2: Verify Condition (\ref{eq:2}) holds for
$\hat{\vb}_i$.}

For $\hat{\vb}_i\in \Nr_i$ satisfying the condition
(\ref{eq:2}), we need to verify
\[\| \vXcr_{i2}^\trans(\vx_i - \vX_{-i}\hat{\vb}_i) \|_\infty <
np'(0+;\lambda_i,\tau_i)\] on the event $\Er_i$.  Note that
\[\vXcr_{i2}^\trans (\vx_i - \vX_{-i}\hat{\vb}_i) =
\vXcr_{i2}^\trans(\vx_i - \vX_{-i}\vb_i) -
\vXcr_{i2}^\trans(\vX_{-i}\hat{\vb}_{i} - \vX_{-i}\vb_{i}) = \bxi_{i2}
- \vXcr_{i2}^\trans\vXcr_{i1}(\hat{\vb}_{i1} - \vb_{i1}).\] By
Condition 1, $\lVert \rho'(\hat{\vb}_{i1};\tau_i) \rVert_\infty \leq
\rho'(\delta_n;\tau_i)$.  On $\Er_i$, by Assumptions (A4) and (A5) we have
\begin{align*} 
  & \| \vXcr_{i2}^\trans(\vx_i - \vX_{-i}\hat{\vb}_i) \|_\infty\\
  \leq & \|\bxi_{i2}\|_\infty + \|
  \vXcr_{i2}^\trans\vXcr_{i1}(\hat{\vb}_{i1} - \vb_{i1})\|_\infty \\
  \leq & \sigma_in^{1/2+a/2}\sqrt{\log(n)} + \lVert \vXcr_{i2}^\trans
  \vXcr_{i1} (\vXcr_{i1}^\trans\vXcr_{i1})^{-1} \rVert_\infty \left[
    \lVert \bxi_{i1}\rVert_\infty + n \lVert
    p'(\hat{\vb}_{i1};\lambda_i,\tau_i) \rVert_\infty \right]\\
  \leq & \sigma_in^{1/2+a/2+b}\sqrt{\log(n)} +
  Knp'(0+;\lambda_i,\tau_i)\\
  < & np'(0+;\lambda_i,\tau_i)
\end{align*} for sufficiently large $n$. \\

\noindent\underline{Step 3: Prove that $\PP(\Er_i) >
  1-C\exp\{n^a-n^a\log(n)/2\}$.}

Since $\lVert \vx_i \rVert_2 = \sqrt{n}$,
$(\sqrt{n}\sigma_i)^{-1}\xi_{ij} \sim \mathrm{N}(0,1)$. 
We have
\begin{align*} 
  \PP(\Er_i) & \geq 1 - \sum_{j=1}^q \PP\left\{
    (\sqrt{n}\sigma_i)^{-1} \lvert
    \xi_{ij} \rvert > n^{a/2} \sqrt{\log(n)}\right\} \\
  & > 1-Cp \exp\left(-\frac{n^a}{2}\log(n) - \frac{a}{2}\log(n) - \log
    \log(n) \right)\\
  & > 1- C\exp\left\{n^a-n^a\log(n)/2 \right\}.
\end{align*}
The last inequality is due to Assumption (A1).

\subsection{Proof of Corollary~1}

Let $\Er=\bigcap_{i=1}^{p_n}\Er_i$ where $\Er_i$ defined in
(\ref{eqn:eps}).  Therefore $\PP(\Er) \geq 1- \sum_{i=1}^{p_n}(1 -
\PP(\Er_i)) \geq1- C\exp\{2n^a - n^a\log(n)/2\} \rightarrow 1$.

\subsection{Proof of Lemma~3}

Suppose that two vertices $i$ and $j$ are not connected in the
skeleton $\mG_n^u$, but connected in the GGM $\mC_{\mG_n}$. In addition, they are not marginally independent. 
By Lemma 2, there exists at least one vertex $k$ such that $i\rightarrow k
\leftarrow j$. Let $\adj_j(i,\mC_{\mG_n}) = \adj(i,\mC_{\mG_n}) \setminus j$. 
Let $\ch_{\mG_n}(i)$ and $\de_{\mG_n}(i)$ be the sets of children and 
descendants of $i$ in $\mG_n$. Let $\ch_{\mG_n}(i,j) = \ch_{\mG_n}(i) \cap
\ch_{\mG_n}(j)$, i.e., the common children of $i$ and $j$. 
Let $\pi_i = \adj_j(i,\mC_{\mG_n}) \setminus \left[
\bigcup_{v\in \ch_{\mG_n}(i,j)}\left(\{v\}\bigcup\de_{\mG_n}(v) \right)\right]$ and
$\pi_j = \adj_i(j,\mC_{\mG_n}) \setminus \left[ \bigcup_{v\in
\ch_{\mG_n}(i,j)}\left(\{v\}\bigcup\de_{\mG_n}(v)
\right)\right]$. We show that $i$ and $j$ is d-separated by $\pi_i
\bigcup \pi_j$ and $\pi_i \bigcup \pi_j\in \Pi_{i,j}$.

In order to show that $i\perpD j | (\pi_i \bigcup \pi_j)$, we consider a
sequence of vertices $k_1,\ldots,k_m$ for $m\geq 1$ of chains such
that \\ \indent (Chain 1) $i\rightarrow k_1-\ldots- k_m\leftarrow j$,\\
\indent (Chain 2) $i\rightarrow k_1-\ldots- k_m\rightarrow j$,\\
\indent (Chain 3) $i\leftarrow k_1-\ldots- k_m\leftarrow j$,\\ 
\indent (Chain 4) $i\leftarrow k_1-\ldots- k_m\rightarrow j$. 
 
These four cases cover all possible chains connecting $i$ and $j$ while we allow $k_1$ and $k_m$ to be the same. It suffices to show that
$\pi_i \bigcup \pi_j$ blocks all the four types of chains between $i$
and $j$. For the (Chain 2), a set including the arrow emitting vertex
$k_m$ d-separates $i$ and $j$ by Definition 1 on d-separation. Since
$k_m\in \adj_i(j,\mC_{\mG_n})$ and $k_m\notin \ch_{\mG_n}(j)$ because
of no loop restriction, $k_m\in \pi_i\bigcup \pi_j$. Similarly for the (Chain 3),
since the arrow emitting vertex $k_1\in \adj_j(i,\mC_{\mG_n})$ but
$k_1\notin \ch_{\mG_n}(i)$, $k_1\in \pi_i\bigcup \pi_j$. The (Chain 4)
also blocked by either arrow-emitting vertices $k_1$ or $k_2$ included
in $\pi_i\bigcup \pi_j$. In the (Chain 1), there must be at least one
collider. If $m=1$, then $k_m$ is a common child so that it is
excluded from $\pi_i\bigcup\pi_j$. If $m=2$, the possible chains are
$i\rightarrow k_1\rightarrow k_2 \leftarrow j$ or $i\rightarrow
k_1\leftarrow k_2 \leftarrow j$ and both chains have one arrow
emitting vertex, $k_1$ or $k_2$ in $\pi_i\bigcup \pi_j$. Now we
suppose that there are at least three vertices, $m>2$. If at least one
of $k_1$ and $k_m$ is not a collider, there exists a arrow emitting
vertex in $\pi_i\bigcup \pi_j$. If both $k_1$ and $k_m$ are colliders,
the (Chain 1) is $i\rightarrow k_1 \leftarrow
k_2-\ldots-k_{m-1}\rightarrow k_m \leftarrow j$. Since the arrow
emitting vertices $k_2$ and $k_{m-1}$ are not in
$\ch_{\mG_n}(i)\bigcap \ch_{\mG_n}(j)$ but in
$\adj_j(i,\mC_{\mG_n})\bigcup \adj_i(j,\mC_{\mG_n})$, those are in
$\pi_i\bigcup \pi_j$. Therefore, $\pi_i\bigcup\pi_j$ blocks all chains
between $i$ and $j$.

Next we need to prove $\pi_i \bigcup \pi_j \in \Pi_{i,j}$. Let
$V_{n,-i,-j} = V_n\setminus\{i,j\}$. Since $\pi_i \bigcup \pi_j =
\left[\adj_j(i,\mC_{\mG_n}) \bigcup \adj_i(j,\mC_{\mG_n})\right]
\setminus \left[ \bigcup_{v\in \ch_{\mG_n}(i,j) }\left(\{v\}\bigcup\de_{\mG_n}(v) \right)\right]$, it
is obvious that
\begin{equation*} 
\bigcup_{v\in \ch_{\mG_n}(i,j)}\left(\{v\}\bigcup \de_{\mG_n}(v)\right) 
\subseteq \bigcup_{v\in \adj(i,j,\mC_{\mG_n})}\Con\big(v,\mC_{\mG_n}(V_{n,-i,-j})\big),
\label{rel1}
 \end{equation*} 
and thus $\pi_i \bigcup \pi_j \in \Pi_{i,j}$.

\subsection{Lemma~6} We state Lemma~6 which is used to prove
Theorem~2. This lemma is essentially the same as Lemma 3 in
\cite{kalisch2007estimating}. The proof is therefore skipped.\\

\noindent Lemma 6: Let $g(\rho) = 0.5 \log((1+\rho)/(1 -
\rho))$. Denote by $\hat{z}_{i,j|\Kr} =
g\left(\hat{\rho}_{i,j|\Kr}\right)$ and by ${z}_{i,j|\Kr} =
g\left({\rho}_{i,j|\Kr}\right)$ where $\Kr \subseteq
\adj(i,\mC_{\mG_n})\bigcup \adj(j,\mC_{\mG_n})$. Assume the
distribution of $X= (X_1, X_2, ..., X_p)^\trans$ is multivariate
Gaussian and $\sup_{i,j,\Kr} \left| \rho_{i, j|\Kr}\right| \leq M < 1$ (the second part of
Assumption (A6)). Then, for any $0 < \gamma < 2$,
  \begin{eqnarray*} \sup_{i,j,\Kr}\bP\left( \left| \hat{z}_{i,j|\Kr} -
{z}_{i,j|\Kr}\right| > \gamma \right) \leq O(n - \nu_i-\nu_j) \left[
\exp\left\{ -(C_1 + C_2) (n - \nu_i-\nu_j - 4) \right\} \right],
  \end{eqnarray*} where $\nu_i = |\adj(i,\mC_{\mG_n})|$ and $C1$ and $C_2$ are two positive constants. 
  More specifically,
$$  C_1 = \log\left[\frac{ 4 + (\gamma l)^2 }{ 4 - (\gamma l)^2 } \right],
\ \ \ C_2 = \log\left[\frac{ 16 + (1 - M)^2}{ 16 - ( 1 - M)^2 }
\right],$$ where $l = 1 - (1+M)^2/4$.

\subsection{Proof of Theorem~2} For an edge $i - j\in F_n$ of
$\mC_{\mG_n}$, define $\Kr$ to be any set in $\boldsymbol{\Pi}_{i,j}$ of
(\ref{eq:pi}) with $|\Kr|<n-3$. Let $\nu_i = |\adj(i,\mC_{\mG_n})|$
for all $i\in V_n$. From Lemma 5 in the Supplementary Materials, if
$\gamma\rightarrow 0$, $C_1 \sim (\gamma l)^2/2 \rightarrow 0$. In
contrast, $C_2$ is a constant. Therefore the term $\exp\{-C_2 (n -
\nu_i-\nu_j -4) \}$ is negligible, and thus
\begin{align*} \sup_{i,j,\Kr}\PP\left( \left\lvert \hat{z}_{i,j|\Kr}-
{z}_{i,j|\Kr}\right\rvert > \gamma \right) &\leq O(n - \nu_i-\nu_j)
\exp\left\{ - (\gamma l)^2(n -\nu_i-\nu_j - 4)/2 \right\} \\ &\leq O(n
- \nu_i-\nu_j) \exp\left\{ - C_3 (n - \nu_i-\nu_j) \gamma^2 \right\},
\end{align*} where $C_3$ is a constant.

Denote by $E_{i,j|\Kr}$ the event ``an error occurred when testing
partial correlation for zero at nodes $i$, $j$ with conditional set
$\Kr$''. An error can be a type I error or a type II error, denoted by
$E_{i,j|\Kr}^{I}$ and $E_{i,j|\Kr}^{II}$, respectively. Therefore
$E_{i,j|\Kr} = E_{i,j|\Kr}^{I} \bigcup E_{i,j|\Kr}^{II}$, and
\begin{align*} E_{i,j|\Kr}^{I}: & \ \sqrt{n - |\Kr| - 3} \left|
\hat{z}_{i,j|\Kr}\right| > \Phi^{-1} (1 - \alpha/2) \textrm{ and }
{z}_{i,j|\Kr} = 0,\\ E_{i,j|\Kr}^{II}: & \ \sqrt{n - |\Kr| - 3} \left|
\hat{z}_{i,j|\Kr}\right| \leq \Phi^{-1} (1 - \alpha/2) \textrm{ and }
{z}_{i,j|\Kr} \neq 0.
\end{align*}

Choose $\alpha = \alpha_n = 2(1 - \Phi(\sqrt{n}c_n/2))$, where $c_n$
is defined in Assumption (A3). Then
\begin{align*} \sup_{i,j,\Kr} \bP (E_{i,j|\Kr}^{I}) &= \sup_{i,j,\Kr}
\PP \left[ \left| \hat{z}_{i,j|\Kr}- {z}_{i,j|\Kr}\right| > \sqrt{n/(n
- |\Kr| - 3)} c_n /2 \right]\\ & \leq O(n - \nu_i-\nu_j)
\exp\left[-C_4(n - \nu_i-\nu_j)c_n^2\right],
\end{align*} for some constant $C_4$. With the same choice of
$\alpha$,
\begin{align*} \sup_{i,j,\Kr} \PP (E_{i,j|\Kr}^{II}) &= \sup_{i,j,\Kr}
\PP \left[ \left| \hat{z}_{i,j|\Kr}\right| \leq \sqrt{n/(n -
\lvert\Kr\rvert - 3)} c_n /2 \right]\\ & \leq \sup_{i,j,\Kr} \PP
\left[ \left| \hat{z}_{i,j|\Kr} - {z}_{i,j|\Kr}\right| > c_n \left(1 -
\sqrt{n/(n - \lvert\Kr\rvert - 3)}/2 \right)\right]\\ & \leq O(n -
\nu_i-\nu_j) \exp\left[-C_5(n - \nu_i - \nu_j)c_n^2\right],
\end{align*} for some constant $C_5$.
\begin{align} & \PP (\textrm{an error occurs in the step 2 of PenPC
algorithm}) \nonumber\\ & \leq \sum_{(i,j)\in F_n} 2^{\nu_i + \nu_j}
O((n-\nu_i-\nu_j))\exp\{-C_6(n-\nu_i-\nu_j)c_n^2\}\nonumber\\ & \leq
O\left[\sum_{i=1}^{p_n}\sum_{j\in
\adj(i,\mC_{\mG_n})}n2^{2q_n}\exp\left\{-C_6(n-2q_n)c_n^2\right\}\right]\nonumber\\
& \leq
O\left[np_nq_n\exp\left\{2q_n-C_6(n-2q_n)c_n^2\right\}\right]\nonumber\\
& \leq
O\left[np_nq_n\exp\left\{-C_6n^{1-2d_1}+C_7q_n\right\}\right]\label{ineq1}\\
& \leq O\left[n^{b+1}
\exp\left\{-C_6n^{1-2d_1}+n^a+C_7n^b\right\}\right]\nonumber
\end{align} for a positive constant $C_6$ and $C_7$. This probability
converges to zero as $n\rightarrow \infty$ when
$0<d_1<\min\left(\frac{1-a}{2},\frac{1-b}{2}\right)$.

\subsection{Proof of Corollary~2}

From Corollary~1 and Theorem~2,
\begin{align*} &\PP(\textrm{an error occurs in the PenPC algorithm})\\
& = \PP(\hat{\mC}_{\mG_n}(\btheta)\neq \mC_{\mG_n}) +
\PP(\hat{\mG}_n^u(\alpha_n)\neq\mG_n^u)\\ & =
O\left(\exp\{2n^a-n^a\log(n)\}\right) +
O\left(\exp\{-Cn^{1-2d_1}\}\right)\\ & = O\left(
\exp\{-Cn^{1-2d_1}\}\right)
\end{align*} for $d_1<min((1-a)/2,(1-b)/2)$.

\end{document}